\documentclass[a4paper,10pt]{article}
\usepackage{graphicx}
\usepackage{amsmath}
\usepackage{citesort}
\usepackage[sort&compress,numbers]{natbib}
\usepackage{amsfonts}
\usepackage{amssymb}
\usepackage{amsmath}
\usepackage{latexsym}
\usepackage{color}
\usepackage{ntheorem}
\usepackage{cancel}
\usepackage[normalem]{ulem}
\usepackage{braket}
\usepackage[colorlinks=true]{hyperref}

\newcounter{mnotecount}[section]

\renewcommand{\themnotecount}{\thesection.\arabic{mnotecount}}
\newcommand{\mnote}[1]
{\protect{\stepcounter{mnotecount}}$^{\mbox{\footnotesize
$
\bullet$\themnotecount}}$ \marginpar{
\raggedright\tiny\em
$\!\!\!\!\!\!\,\bullet$\themnotecount: #1} }

\newcommand{\tim}[1]{\mnote{{\bf tim:}#1}}

\newtheorem{theorem}{\sc  Theorem\rm}[section]

\newtheorem{corollary}[theorem]{\sc  Corollary\rm}

\newtheorem{lemma}[theorem]{\sc Lemma\rm}

\newtheorem{proposition}[theorem]{\sc Proposition\rm}

\newtheorem{remark}[theorem]{\sc Remark\rm}

\newcommand{\jlcax}[1]{}
\newcommand{\eean}{\nonumber\end{eqnarray}}

\newcommand{\kk}[1]{}

\newcommand{\beq}{\begin{equation}}

\newcommand{\FS}       
                  {F}

\newcommand{\HS} 
       {H_{\mbox{\scriptsize volume}}}

{\ptc{this should be removed in the oberwolfach version}}%

\newcommand{\eeal}[1]{\label{#1}\end{eqnarray}}
\newcommand{\bed}{\begin{deqarr}}
\newcommand{\eed}{\end{deqarr}}
\newcommand{\bedl}[1]{\begin{deqarr}\label{#1}}
\newcommand{\eedl}[2]{\arrlabel{#1}\label{#2}\end{deqarr}}

\newcommand{\bel}[1]{\begin{equation}\label{#1}}
\newcommand{\bea}{\begin{eqnarray}}
\newcommand{\bean}{\begin{eqnarray}\nonumber}
\newcommand{\beal}[1]{\begin{eqnarray}\label{#1}}
\newcommand{\eea}{\end{eqnarray}}

\newcommand{\qed}{\hfill $\Box$ \medskip}

\newcommand{\be}{\begin{equation}}
\newcommand{\eeq}{\end{equation}}
\newcommand{\ee}{\end{equation}}
\newcommand{\beqa}{\begin{eqnarray}}
\newcommand{\eeqa}{\end{eqnarray}}
\newcommand{\beqan}{\begin{eqnarray*}}
\newcommand{\eeqan}{\end{eqnarray*}}
\newcommand{\ba}{\begin{array}}
\newcommand{\ea}{\end{array}}

\newcommand{\mcM}{{\mycal M}}

\newcommand{\scri}{{\mycal I}}%

\newcommand{\warn}[1]
{\protect{\stepcounter{mnotecount}}$^{\mbox{\footnotesize
$
\bullet$\themnotecount}}$ \marginpar{
\raggedright\tiny\em
$\!\!\!\!\!\!\,\bullet$\themnotecount: {\bf Warning:} #1} }

\newcommand{\eq}[1]{(\ref{#1})}

\newcommand{\ptc}[1]{\mnote{{\bf ptc:}#1}}

\newcommand{\mcL}{{\mycal L}}

\newcommand{\beqar}{\begin{deqarr}}
\newcommand{\eeqar}{\end{deqarr}}

\newcommand{\beaa}{\begin{eqnarray*}}
\newcommand{\eeaa}{\end{eqnarray*}}

\newcommand{\tr}{\mbox{tr}}

\DeclareFontFamily{OT1}{rsfs}{}
\DeclareFontShape{OT1}{rsfs}{m}{n}{ <-7> rsfs5 <7-10> rsfs7 <10-> rsfs10}{}
\DeclareMathAlphabet{\mycal}{OT1}{rsfs}{m}{n}

{\catcode `\@=11 \global\let\AddToReset=\@addtoreset}
\AddToReset{equation}{section}

{\catcode `\@=11 \global\let\AddToReset=\@addtoreset}
\AddToReset{figure}{section}

{\catcode `\@=11 \global\let\AddToReset=\@addtoreset}
\AddToReset{table}{section}

\begin{document}

\title{Algorithmic characterization results for the Kerr-NUT-(A)dS space-time. I.\\ A space-time approach%
\thanks{Preprint UWThPh-2016-21. }
}
\author{Tim-Torben Paetz%
\thanks{E-mail:  Tim-Torben.Paetz@univie.ac.at}  \vspace{0.5em}\\  \textit{Gravitational Physics, University of Vienna}  \\ \textit{Boltzmanngasse 5, 1090 Vienna, Austria }
}

\maketitle

\vspace{-0.2em}

\begin{abstract}
We provide an algorithm to check whether a given vacuum space-time $(\mcM,g)$ admits a Killing vector field w.r.t.\ which the  Mars-Simon tensor vanishes. In particular, we obtain an algorithmic procedure to check whether $(\mcM,g)$   is
 locally isometric to a member of the Kerr-NUT-(A)dS family. A particular emphasis will be devoted to the Kerr-(A)dS case.
\end{abstract}


\tableofcontents

\section{Introduction}

One of the most important families of exact solutions to Einstein's vacuum field equations is provided by the Kerr space-time:
It is expected that due to  radiation black holes, which are formed by the gravitational collapse of a star, eventually settle down to an asymptotically flat, stationary vacuum black hole solution.  Kerr black holes are conjectured to be the only solutions with all these properties and therefore, if true, would characterize the asymptotic state of a large class of  evolution processes.

Classical results by Hawking, Carter and Robinson (cf.\ \cite{cc,IK2} for an overview) state that this conjecture is true under the
additional and unfortunately quite unphysical hypothesis of
real analyticity of the space-time.
In more recent results \cite{IK, aik1, aik2} this unwanted analyticity assumption is  dropped. However, certain additional
assumptions still need to be imposed  to end up with a black hole uniqueness result, so that the original conjecture is still open.
In any case,  these somewhat weaker uniqueness results already underline the distinguished role of the Kerr family in general relativity.

Usually the Kerr space-time, or,  more general, the  Kerr-NUT-(A)dS space-time, is given in Boyer-Lindquist-type coordinates,
which are adapted to the stationary and axial symmetries of this family.
In some situations, though, such as e.g.\ in certain numerical applications, one may have to deal with coordinate systems which
do not reflect these symmetries.
In such cases it is convenient to have a  gauge-invariant characterization at hand, for instance to check whether a  metric given in completely arbitrary coordinates belongs to that family.
Here, let us mention different approaches which achieve that  (compare also \cite{bk}):
\begin{enumerate}
\item[(i)] The first one relies on the \emph{Mars-Simon tensor (MST)}, cf.\ \cite{IK}, a space-time analog of the \emph{Simon tensor} \cite{simon,perjes}: Whenever a vacuum space-time with a Killing vector field (KVF) has been given, one can define this tensor, which comes along with all the algebraic symmetries of the Weyl tensor. The vanishing
of this tensor, supplemented by certain additional conditions, provides a characterization of Kerr \cite{mars,mars2}, Kerr-NUT \cite{mars3}, and   Kerr-NUT-(A)dS \cite{mars_senovilla}, respectively.
\item[(ii)]  Another  approach is based on so-called  hidden symmetries:   The existence of a closed, non-degenerate
\emph{conformal Killing-Yano tensor (CKYT)}  implies that the underlying  vacuum space-time belongs to the  Kerr-NUT-(A)dS family \cite{houri,krtous}.
\item[(iii)]
The Kerr space-time can be further  characterized among vacuum space-times with vanishing cosmological constant in terms of the existence of a \emph{Killing spinor} whose associated KVF is real \cite{bk, bk2, cole} (supplemented by certain additional conditions).
\item[(iv)] Finally,  a characterization of the Kerr(-NUT)-family in terms of \emph{concomitants of the Weyl tensor},
which are objects constructed merely from tensorial operations on the Weyl tensor and its covariant derivatives, has been given in \cite{fs}.
\end{enumerate}

These results contribute to a more satisfactory understanding of the Kerr-NUT-(A)dS family from a space-time perspective.
In a sense, they generalize a corresponding result for the Minkowski space-time, which can locally be characterized among vacuum space-times by the vanishing
of the conformal Weyl tensor.
However, there is one decisive  difference: The Weyl tensor can be computed straightforwardly from the line element just by differentiation.

However,
in order to apply the above characterization results (i) and (ii), one first of all needs to check whether a given vacuum space-time   admits a KVF or CKYT,  respectively. This is a delicate  issue which requires to solve partial differential equations.
Only once this step has been accomplished and a solution has been found, it becomes straightforward to check whether the remaining hypotheses are fulfilled.
In the case where  they  do not hold, though, there might exist another KVF or CKYT for which they do hold.

The characterization result (iv), which relies on concomitants of the Weyl tensor,  is \emph{algorithmic}: It provides a way to check whether a given vacuum space-time is Kerr(-NUT) just by performing differentiations and algebraic operations, without any need of solving differential equations.
A Killing spinor is defined by a differential equation.
A classical result (cf. \cite{bk}), though,  says that  a vacuum space-time admits a Killing spinor if and only if  it is of Petrov type D, N or O, and Killing spinor candidates can be computed from the Weyl spinor, whence the Killing spinor  approach (iii)  is  algorithmic, as well.

In this work we focus attention on the characterization results (i) which are based on the MST $\mathcal{S}_{\mu\nu\sigma\rho}$ as defined in \eq{dfn_mars-simon} below.
Our aim is to modify this approach to permit an  algorithmic characterization.
  We will show that
the requirement of a vanishing
MST imposes  restrictions on the associated KVF, which turn out to be so strong, that for a given vacuum space-time with cosmological constant $\Lambda$ there remains
at most one candidate field $X$ (up to rescaling).
This candidate field can  be computed \emph{algebraically} from concomitants of the Weyl tensor by, roughly speaking,
solving $\mathcal{S}_{\mu\nu\sigma\rho}=0$ for the KVF.
In fact, it turns out that whenever this equation admits a solution $X$, it has automatically the desired properties, i.e.\
defines a (possibly complex) KVF such that  its  associated  MST vanishes.
The solvability of $\mathcal{S}_{\mu\nu\sigma\rho}=0$ for $X$ can be characterized by the vanishing of a certain tensor which is constructed
algebraically from the Weyl tensor.
It is closely related to a certain reformulation of  the Petrov type D condition, which, though, in itself is not sufficient for this, but needs to be supplemented
by an additional condition.
 This is our first main result, cf.\ Theorem~\ref{int_main_result}. For the convenience of the reader we provide a somewhat shortened version already here
 ($\mathcal{C_{\alpha\beta\gamma\delta}}$ denotes the self-dual Weyl tensor and we set $\mathcal{I}_{\mu\nu\sigma\rho}:=\frac{1}{4} (g_{\mu\sigma}g_{\nu\rho} -g_{\mu\rho}g_{\nu\sigma} + i\epsilon_{\mu\nu\sigma\rho} )$):

\begin{theorem}
\label{int_main_result_intro}
Let $(\mcM,g)$ be a  smooth $(3+1)$-dimensional $\Lambda$-vacuum space-time
containing an open dense set on which
\begin{equation}
\mathcal{C}^2 \,\ne\, 0\;, \quad  \mathcal{C}^2 - \frac{32}{3}\Lambda^2\,\ne\, 0\;,
\quad   \mathcal{C}^2 -\frac{8}{3} \Lambda^2 \,\ne\, 0
\;.
\label{all_ineq_C2}
\end{equation}
Then, if and only if
\begin{itemize}
\item  the self-dual Weyl tensor corrected by a trace-term,
$\mathcal{W}_{\alpha\beta\mu\nu} := \mathcal{C}_{\alpha\beta\mu\nu}
\pm \sqrt{\frac{\mathcal{C}^2}{6}}\,\mathcal{I}_{\alpha\beta\mu\nu}$, satisfies $\mathcal{W}=\omega\otimes\omega$ for some two-form $\omega$,
\end{itemize}
or, equivalently,
\begin{itemize}
\item
the Weyl field
$\mathfrak{C}_{\alpha\beta\gamma\delta}\equiv  \mathcal{C}_{\alpha\beta\mu\nu} \mathcal{C}^{\mu\nu}{}_{\gamma\delta}
\mp \sqrt{\frac{\mathcal{C}^2}{6}}\,\mathcal{C}_{\alpha\beta\gamma\delta}
-
\frac{\mathcal{C}^2}{3}\,\mathcal{I}_{\alpha\beta\gamma\delta} $
vanishes
\end{itemize}
(the conditions will be satisfied for at most one sign $\mp$),
there exists an (up to rescaling) unique, non-trivial, possibly complex KVF $X$ such that the associated MST vanishes.
\end{theorem}

Based on this result algorithmic characterizations of the  Kerr-NUT-(A)dS family are obtained by employing the results
in \cite{mars_senovilla}.
Indeed,  a check of  the  remaining hypotheses which are needed to apply the result in (i) requires merely algebraic manipulations and differentiation and is therefore just a matter of computation.
Particular attention will be devoted to the Kerr-(A)dS case
where the NUT-parameter vanishes.
These results provide an alternative algorithm  compared to those  in \cite{bk, fs}
 for $\Lambda=0$, which, furthermore,  extends
to the $\Lambda\ne 0$-case. This is our second main result, cf.\ Theorem~\ref{thm_main_result}, a somewhat 
shortened version of which reads:

\begin{theorem}
\label{thm_main_result_intro}
Let  $(\mcM, g)$ be a smooth
non-maximally symmetric
$(3+1)$-dim.\ space-time which satisfies  Einstein's vacuum field equations
with cosmological constant $\Lambda$. Then,  the space-time $(\mcM, g)$ is locally isometric to a member of the Kerr-(A)dS family  if and only if
\begin{enumerate}
\item[(i)] $\mathcal{C}^2\ne 0$ everywhere,
\item[(ii)] $\mathfrak{C}_{\alpha\beta\gamma\delta}\,\equiv \, \mathcal{C}_{\alpha\beta\mu\nu} \mathcal{C}^{\mu\nu}{}_{\gamma\delta}
\mp\sqrt{\frac{\mathcal{C}^2}{6}}\,\mathcal{C}_{\alpha\beta\gamma\delta}
-
\frac{\mathcal{C}^2}{3}\,\mathcal{I}_{\alpha\beta\gamma\delta}= 0 $ (a solution exists at most for either $+$ or $-$),
\item[(iii)]
$\mathcal{C}^2 \ne \frac{32}{3}\,\Lambda^2$, and $\mathcal{C}^2\ne \frac{8}{3} \Lambda^2 $,
\item[(iv)] $
 \mathrm{Re}\Big[(\mathcal{C}^2)^{-7/6}
\Big(\mathcal{C}^{\alpha\beta}{}_{\mu\nu} \mathcal{W}^{\mu\nu}
\pm \sqrt{ \frac{\mathcal{C}^2}{6}}\,\mathcal{W}^{\alpha\beta}\Big) \nabla_{\beta} \mathcal{C}^2
\Big]=0
$
 for some
(and then any)  self-dual 2-tensor   $\mathcal{W}_{\mu\nu}$ which satisfies $| \mathcal{C}^{\alpha\beta}{}_{\mu\nu} \mathcal{W}^{\mu\nu}
\pm\sqrt{ \frac{\mathcal{C}^2}{6}}\,\mathcal{W}^{\alpha\beta}|^2 =1$,
\item[(v)] $\mathrm{grad}(\mathrm{Re} [ (\mathcal{C}^2)^{-1/6}])$ is not identically zero, and
\item[(vi)] the constants  $c$ and $k$,
 given in terms of $\mathcal{C}^2$ by
\eq{c_KdS}-\eq{k_KdS} (and \eq{gauge_varkappa}),
 satisfy, depending on the sign of the cosmological constant, \eq{second_condition1}-\eq{second_condition3}, respectively.
\end{enumerate}
\end{theorem}

Another issue,  which complements these kind of problems, is to derive analog  results from the point of view of an initial value  problem.
One would like know whether a given set of Cauchy data (or characteristic initial data)  generates
a development which is isometric to a portion of a Kerr or, again more general,  a Kerr-NUT-(A)dS space-time.
The derivation of a corresponding algorithm, based on the results of this work, will be the content of part II \cite{kerr_part2} of this paper.

The article is organized as follows:
In Section~\ref{section_preliminaries} we define the MST. Moreover, we recall some space-time characterization results
for  Kerr-NUT-(A)dS on which our analysis is based.
In Section~\ref{section_alg_charact} we obtain our first main result, Theorem~\ref{int_main_result}.
To prove it,  we will start  in Section~\ref{sec_function_Q} with a careful discussion concerning  the choice of the function $Q$ which appears in the definition of the MST.
In doing so, we will clarify the relation between different definitions of $Q$, in the setting where the MST
vanishes. Proceeding in this way we also find some relations which are crucial in view of the algorithmic approach we are aiming at.

We will  show in Sections~\ref{sec_petrov}-\ref{sec_solv} that, in $\Lambda$-vacuum, the equation of a vanishing MST can, under certain circumstances, be solved
 for the KVF $X$, which then can be determined algebraically.
As  indicated above, this is possible because the  condition, that the KVF is such that the associated MST vanishes, is so restrictive that there  remains just one candidate field (up to rescaling).
Proceeding this way we obtain in Sections~\ref{section_alg_charact}-\ref{sec_alt}  an algorithmic characterization of all $\Lambda$-vacuum space-times
which admit a (possibly complex) KVF such that the associated  MST vanishes.
For this step an appropriate choice of the function $Q$ turns out to be  decisive.

The considerations in Section~\ref{sec_KVF_exist} will show that the so-obtained candidate field is automatically a KVF
whose associated MST vanishes. It  is the only KVF with this property (cf.\  Section~\ref{sec_uniqueness}).
The only ``obstruction'' for the existence of a KVF with vanishing MST is therefore
the solvability issue which is solved by the above mentioned Theorem~\ref{int_main_result}.
A straightforward application  of the available characterization results recalled in Section~\ref{section_preliminaries}  then  provides
an algorithmic characterization of the Kerr-NUT-(A)dS-family in Sections~\ref{sec_van_real_KVF}-\ref{sec_NUT}.
Specializing this to the case where the NUT-parameter vanishes, some expressions can be made somewhat more explicit.
This is accomplished in our second main result in Section~\ref{sec_Kerr_AdS},  Theorem~\ref{thm_main_result}.

%



In this work we will deal with smooth $(3+1)$-dimensional space-times which satisfy  Einstein's vacuum field equations
with cosmological constant $\Lambda\in\mathbb{R}$.
The results remain true if only finite differentiability is assumed. We leave it to the reader to work  out the precise requirements which are needed
for this.

\section{Preliminaries}
\label{section_preliminaries}

\subsection{Self-dual tensors}
\label{sect_self-dual}

In this  section we will present  some useful formulas  for  self-dual tensors (cf.\ e.g.\ \cite{IK}).
Let $(\mcM, g)$ be a smooth $(3+1)$-dimensional space-time.
A   two-form $\mathcal{W}_{\alpha\beta}$ is called \emph{self-dual} if it holds that
\begin{equation}
  \mathcal{W}_{\alpha\beta} \,=\, i \mathcal{W}^{\star}_{\alpha\beta}
\,,
\end{equation}
where
\begin{equation}
\mathcal{W}^{\star}_{\alpha\beta} \,:=\, \frac{1}{2}\epsilon_{\alpha\beta}{}^{\mu\nu}\mathcal{W}_{\mu\nu}
\end{equation}
denotes the \emph{Hodge dual}.

Any two self-dual two-forms $\mathcal{V}_{\alpha\beta}$ and  $\mathcal{W}_{\alpha\beta}$ satisfy
\begin{eqnarray}
\mathcal{V}_{(\mu|\alpha|}\mathcal{W}_{\nu)}{}^{\alpha} &=& \frac{1}{4} g_{\mu\nu}\mathcal{V}_{\alpha\beta}\mathcal{W}^{\alpha\beta}
\,,
\\
\mathcal{W}_{\mu\alpha}\mathcal{W}_{\nu}{}^{\alpha} &=&\frac{1}{4}\mathcal{W}^2g_{\mu\nu}\,,
\end{eqnarray}
where
\begin{equation}
\mathcal{W}^2 \,:=\, \mathcal{W} _{\alpha\beta}\mathcal{W}^{\alpha\beta}
\,.
\end{equation}

A tensor $\mathcal{U}_{\alpha\beta\mu\nu}$ is called a \emph{double two-form} if
\begin{equation}
   \mathcal{U}_{\alpha\beta\mu\nu}\,=\,- \mathcal{U}_{\beta\alpha\mu\nu} \,=\, -\mathcal{U}_{\alpha\beta\nu\mu}
\,.
\label{part_skew}
\end{equation}
A double two-form $\mathcal{U}_{\alpha\beta\mu\nu}$ is called  \emph{left  (right) self-dual} if
\begin{equation}
\mathcal{U}_{\alpha\beta\mu\nu} \,=\, i {}^{\star}\mathcal{U}_{\alpha\beta\mu\nu}
\quad
(
\mathcal{U}_{\alpha\beta\mu\nu} \,=\, i \mathcal{U}^{\star}_{\alpha\beta\mu\nu}
)
\label{self_dual4}
\end{equation}
where
\begin{equation}
{}^{\star}\mathcal{U}_{\alpha\beta\mu\nu}\,:=\, \frac{1}{2}\epsilon_{\alpha\beta}{}^{\gamma\delta}\mathcal{U}_{\gamma\delta\mu\nu}
\quad
(
\mathcal{U}^{\star}_{\alpha\beta\mu\nu}\,:=\, \frac{1}{2}\epsilon_{\mu\nu}{}^{\gamma\delta}\mathcal{U}_{\alpha\beta\gamma\delta}
)
\end{equation}
denotes the \emph{left (right) Hodge dual}.

Let $\mathcal{W}_{\alpha\beta}$ be a self-dual two-form, and  $\mathcal{U}_{\alpha\beta\mu\nu}$, $\mathcal{V}_{\alpha\beta\mu\nu}$ be right self-dual 4-tensors, then
\begin{eqnarray}
\mathcal{W}_{(\mu}{}^{\gamma}\mathcal{U}_{|\alpha\beta|\nu)\gamma} &=& \frac{1}{4} g_{\mu\nu}\mathcal{W}^{\gamma\delta}
\mathcal{U}_{\alpha\beta\gamma\delta}
\,,
\\
 \mathcal{U}_{\mu\alpha(\gamma}{}^{\sigma} \mathcal{V}_{|\nu\beta|\delta)\sigma} &=& \frac{1}{4} g_{\gamma\delta} \mathcal{U}_{\mu\alpha}{}^{\kappa\varkappa} \mathcal{V}_{\nu\beta\kappa\varkappa}
\label{weyl_rel1}
\;.
\end{eqnarray}

A double two-form  is called a \emph{Weyl field} if it has all the algebraic symmetries of the Weyl tensor, i.e.\
if, in addition to \eq{part_skew}, the following relations are fulfilled,
\begin{eqnarray}
\mathcal{U}_{\alpha\beta\mu\nu} \,=\, \mathcal{U}_{\mu\nu\alpha\beta}\,,\quad
\mathcal{U}_{\alpha[\beta\mu\nu]}\,=\, 0\,, \quad  g^{\beta\nu}\mathcal{U}_{\alpha\beta\mu\nu}\,=\,0
\,.
\label{Weyl_field}
\end{eqnarray}
If  $\mathcal{U}_{\alpha\beta\mu\nu}$ is a Weyl field, ${}^{\star}\mathcal{U}_{\alpha\beta\mu\nu}$ and $\mathcal{U}^{\star}_{\alpha\beta\mu\nu}$ are also Weyl fields, and both Hodge duals coincide. Such a left (and then also right) self-dual tensor is simply called \emph{self-dual}.

Finally, we set
\begin{equation}
\mathcal{I}_{\mu\nu\sigma\rho} \,:=\, \frac{1}{4} (g_{\mu\sigma}g_{\nu\rho} -g_{\mu\rho}g_{\nu\sigma} + i\epsilon_{\mu\nu\sigma\rho} )
\label{dfn_I}
\,,
\end{equation}
which is a (left and right) self-dual rank-4 tensor. Let $\mathcal{W}_{\alpha\beta}$ be a self-dual two-form,
then
\begin{eqnarray}
\mathcal{I}_{\alpha\beta\mu\nu}\mathcal{W}^{\mu\nu} &=& \mathcal{W}_{\alpha\beta}
\,,
\end{eqnarray}
i.e.\  $\mathcal{I}_{\mu\nu\sigma\rho}$ may be viewed as a metric in the space of self-dual two-forms.

\subsection{Mars-Simon tensor (MST)}
\label{sect_MST}

\label{sect_MST1}
Let $(\mcM, g)$ be a smooth $(3+1)$-dimensional space-time which admits a Killing vector field (KVF) $X$.
Denote by $C_{\mu\nu\sigma\rho}$ its \emph{conformal Weyl tensor}, while
\begin{equation}
F_{\mu\nu} \,:=\,  \nabla_{\mu}X_{\nu} =  \nabla_{[\mu}X_{\nu]}
\label{rln_X_F}
\end{equation}
denotes the associated \emph{Killing form}.
We then define the  \textit{Mars-Simon tensor (MST)} (compare \cite{IK} where somewhat different conventions are used)
as
\begin{equation}
\mathcal{S}_{\mu\nu\sigma\rho}\,:=\, \mathcal{C}_{\mu\nu\sigma\rho}  + Q\mathcal{Q}_{\mu\nu\sigma\rho}
\,,
\label{dfn_mars-simon}
\end{equation}
where
\begin{eqnarray}
\mathcal{Q}_{\mu\nu\sigma\rho}  &:=& -  \mathcal{F}_{\mu\nu}\mathcal{F}_{\sigma\rho} + \frac{1}{3}\mathcal{F}^2\mathcal{I}_{\mu\nu\sigma\rho}
\,,
\end{eqnarray}
and where
\begin{eqnarray}
 \mathcal{C}_{\mu\nu\sigma\rho} &:=& C_{\mu\nu\sigma\rho} +i C^{\star}_{\mu\nu\sigma\rho}
\,,
\\
\mathcal{F}_{\mu\nu} &:=& F_{\mu\nu} +i F^{\star}_{\mu\nu}
\,,
\label{dfn_calF}
\end{eqnarray}
denote the \emph{self-dual Weyl tensor} and the \emph{self-dual Killing form}, respectively, which satisfy
\begin{equation}
\mathcal{F}_{\mu\nu} \,=\, i\mathcal{F}^{\star}_{\mu\nu}\,, \quad \mathcal{C}_{\mu\nu\sigma\rho} \,=\,
i \mathcal{C}^{\star}_{\mu\nu\sigma\rho}
\,.
\label{self_dual}
\end{equation}
The MST is a Weyl field, i.e.\ it  satisfies \eq{part_skew} and \eq{Weyl_field}.
At this stage $Q:\mcM\rightarrow \mathbb{C}$ is an arbitrary function. A thorough discussion on how to choose $Q$ will be given in Section~\ref{sec_function_Q}.

Let us collect
 some useful formulas satisfied by a self-dual Killing form in any $\Lambda$-vacuum space-time $R_{\mu\nu}=\Lambda g_{\mu\nu}$ \cite{mars_senovilla},
\begin{eqnarray}
\nabla_{\mu}\mathcal{F}_{\alpha\beta} &=& -X^{\nu}\Big(\mathcal{C}_{\mu\nu\alpha\beta} + \frac{4}{3}\Lambda \mathcal{I}_{\mu\nu\alpha\beta}\Big)\;
\label{rln_nablaF}
\\
\nabla_{\mu}\mathcal{F}_{\alpha}{}^{\mu} &=& \Lambda X_{\alpha}\,,
\label{rln_divF}
\\
\nabla_{\mu}\mathcal{F}^2 &=&  -2X^{\nu}\mathcal{F}^{\alpha\beta}\mathcal{C}_{\mu\nu\alpha\beta} +\frac{4}{3}\Lambda \chi_{\mu}
\label{rln_nablaF2}
\,,
\\
\chi^{\alpha}\mathcal{F}_{\alpha\mu} &=& -\frac{1}{2}\mathcal{F}^2 X_{\mu}
\,,
\label{rln_chi}
\end{eqnarray}
where
\begin{equation}
\chi_{\mu} \,:=\, 2X^{\alpha}\mathcal{F}_{\alpha\mu}
\label{ernst_form}
\end{equation}
denotes the \emph{Ernst one-form}.
One straightforwardly checks that in a $\Lambda$-vacuum space-time the Ernst one-form
%
is closed. Thus, at least locally, there exists a scalar field $\chi$, the \emph{Ernst potential}, such that
\begin{equation}
\chi_{\mu} \,=\,\nabla_{\mu}\chi
\,.
\label{ernst_pot}
\end{equation}
At this stage $\chi$ is  defined up to some additive complex constant, the \emph{``$\chi$-constant''}.

The following equations follow from the self-duality of $\mathcal{F}_{\alpha\beta}$, cf.\ Section~\ref{sect_self-dual}, and do
\emph{neither}  rely on the fact that the Killing form arises from a KVF \emph{nor}  on the vacuum equations,
\begin{eqnarray}
\mathcal{F}_{\mu\alpha}\mathcal{F}_{\nu}{}^{\alpha} &=& \frac{1}{4}\mathcal{F}^2g_{\mu\nu}
\,,
\label{self_one}
\label{relation_F2}
\\
 \mathcal{F}_{(\mu}{}^{\sigma}\mathcal{I}_{\nu)\sigma\alpha\beta}
&=&
\frac{1}{4} g_{\mu\nu}\mathcal{F}_{\alpha\beta}
\label{important_relation}
\,,
\\
\mathcal{F}^{\alpha\beta} \mathcal{I}_{\mu\nu\alpha\beta} &=& \mathcal{F}_{\mu\nu}
\label{useful_first}
\,.
\end{eqnarray}

Moreover, the second Bianchi identity together with the $\Lambda$-vacuum Einstein equations imply that the self-dual
Weyl tensor is divergence-free,
\begin{equation}
\nabla^{\alpha }\mathcal{C}_{\alpha\beta\mu\nu} \,=\,0
\,.
\label{useful_last}
\end{equation}

\subsection{Characterization results for space-times with vanishing MST}
\label{sec_results}

Let us start with a characterization result for the Kerr(-NUT) family.

\begin{theorem}[\cite{mars,mars2,mars3}, cf.\  \cite{mars-senovilla-null}]
\label{thm_charact1}
Let $(\mcM,g)$ be a smooth $(3+1)$-dimensional $\Lambda=0$-vacuum space-time which admits a KVF $X$.
Assume that $\mathcal{F}^2$ is not identically zero, and that the MST associated to $X$ vanishes for some function $Q$.
Then, $\mathcal{F}^2\ne 0 $, the Ernst  one-form is exact, and there exists an Ernst potential $\chi$
 and  $0\ne l \in \mathbb{C}$ such that $\mathcal{F}^2=-l\chi^4$ and  $Q=\frac{6}{\chi}$.
Assume further that there exists $q\in\mcM$ with $\mathcal{F}^2|_q\ne 0$ such that $X|_q$ is not orthogonal to the  2-plane generated by the two real null eigenvectors of $\mathcal{F}_{\alpha\beta}|_q$.
\begin{enumerate}
\item[(i)] If, in addition, 
$  -|X|^2-\mathrm{Re}(\chi) >0 $, then $(\mcM,g)$ is locally isometric to a Kerr-NUT space-time.
\item[(ii)] If, in addition, 
$  -|X|^2-\mathrm{Re}(\chi) >0 $ and $l$ is real
and positive,
then $(\mcM,g)$ is locally isometric to a Kerr space-time.
\end{enumerate}
\end{theorem}

In \cite{mars_senovilla, mars-senovilla-null} a complete classification of $\Lambda$-vacuum space-times with vanishing MST
has been given.
In order to state the  main result obtained there,  some auxiliary results are needed.
Following \cite{mars_senovilla}, we define 4 real-valued functions
 $b_1$, $b_2$, $c$ and $k$
(we assume that  $Q\mathcal{F}^2 - 4\Lambda $ is nowhere vanishing),
%
\begin{eqnarray}
 b_2-ib_1 &:=& - \frac{ 36 Q (\mathcal{F}^2)^{5/2} }{(Q\mathcal{F}^2-4\Lambda)^3}
\label{equation_b1b2}
\,,
\\
c &:=& - |X|^2-  \mathrm{Re}\Big(  \frac{6\mathcal{F}^2(Q\mathcal{F}^2+2\Lambda)}{(Q\mathcal{F}^2-4\Lambda)^2} \Big)
\,,
\label{equation_c}
\\
k  &:=&  \Big|\frac{36\mathcal{F}^2}{(Q\mathcal{F}^2-4\Lambda)^2}\Big| \nabla_{\mu}Z\nabla^{\mu}Z -  b_2Z +cZ^2 +\frac{\Lambda}{3} Z^4
\,,
\label{equation_k}
\end{eqnarray}
where
\begin{equation}
 Z = 6\,\mathrm{Re} \Big( \frac{\sqrt{\mathcal{F}^2}}{Q\mathcal{F}^2-4\Lambda}\Big)
\,.
\label{equation_Z}
\end{equation}
The following result is proved in \cite[Theorems 4 \& 6]{mars_senovilla}:
\begin{proposition}
\label{constancy}
Let $(\mcM,g)$ be a  smooth $(3+1)$-dimensional  $\Lambda$-vacuum space-time which admits a KVF $X$ such that the MST vanishes for some function $Q$.
Assume further that the functions $Q\mathcal{F}^2$ and $Q\mathcal{F}^2-4\Lambda$ are not identically zero, and that $\mathrm{Im}\Big(\frac{\sqrt{\mathcal{F}^2}}{Q\mathcal{F}^2-4\Lambda}\Big)$ has non-zero gradient somewhere.
Then:
\begin{enumerate}
\item[(i)]
 $\mathcal{F}^2$ and $Q\mathcal{F}^2-4\Lambda$  are nowhere vanishing,
\item[(ii)]
the Ernst  one-form  $\chi_{\mu}$ is exact, and
\item[(iii)]
$b_1$, $b_2$, $c$ and $k$  are constant.
\end{enumerate}
\end{proposition}

Now, we can state the second characterization result:
\begin{theorem}[\cite{mars_senovilla}]
\label{thm_charact_aux}
Let $(\mcM,g)$ be a smooth $(3+1)$-dimensional $\Lambda$-vacuum space-time which admits a KVF $X$ such that  the associated MST vanishes for some function $Q$.
Assume  that
\begin{equation}
\text{$Q\mathcal{F}^2$ and $Q\mathcal{F}^2-4\Lambda$ are not identically zero,}
\label{charact_cond}
\end{equation}
and that
\begin{equation}
\mathrm{Im}\Big(\frac{\sqrt{\mathcal{F}^2}}{Q\mathcal{F}^2-4\Lambda}\Big) \enspace \text{has non-zero gradient somewhere.}
\label{charact_condB}
\end{equation}
Then
\begin{equation}
\nabla_{\mu}Z\nabla^{\mu} Z \, \geq \, 0
\label{nablaZ_cond}
\;.
\end{equation}
Assume further that the polynomial
\begin{equation}
V(\zeta) \,=\, k + b_2 \zeta -c \zeta^2 - \frac{\Lambda}{3}\zeta^4
\end{equation}
admits two zeros $\zeta_0 \leq \zeta_1$ such that the factor polynomial $\hat V(\zeta) := -  (\zeta-\zeta_0)^{-1}(\zeta-\zeta_1)^{-1} V(\zeta)$
is strictly positive on $[\zeta_0,\zeta_1]$ and that the function $Z$ takes values in $[\zeta_0,\zeta_1]$.
Then $(\mcM, g)$ is locally isometric to a Kerr-NUT-(A)dS space-time with parameters $(\Lambda,m,a,\ell)$, where
\begin{equation}
m\,=\,\frac{b_1}{2v_0^{3/2}}\,, \quad a\,=\, \frac{\zeta_1-\zeta_0}{2\sqrt{v_0}}\,, \quad \ell \,=\, \frac{\zeta_1+\zeta_0}{2\sqrt{v_0}}\,,
\end{equation}
with  $v_0:= \hat V(\frac{\zeta_0+\zeta_1}{2})$.
\end{theorem}


\begin{remark}
\label{remark_nablaZ_cond}
{\rm
The inequality \eq{nablaZ_cond} is equivalent to $V(Z)\geq 0$, which will be used shortly.
}
\end{remark}

We will be particularly interested in a characterization of the Kerr-(A)dS family,
where  the  hypotheses of  Theorem~\ref{thm_charact_aux}  can be made more explicit:
We observe that $\ell=0$ is only possible if%
\footnote{
Conversely, for $\Lambda \ne 0 $, $b_2=0$ does \emph{not} imply $\ell=0$, cf.\  \cite{mpss2} for an example.
}
\begin{equation}
b_2 \,=\, 0
\;.
\end{equation}
In that case the polynomial  $V(\zeta)$
  admits two zeros
$\zeta_0=-\zeta_1 \leq \zeta_1$
 such that the factor polynomial $\hat V$, which then takes the form
\begin{equation}
\hat V \,=\,   \frac{\Lambda}{3}\zeta^2 +c + \frac{\Lambda}{3}\zeta_1^2\;,
\end{equation}
is strictly positive on $[-\zeta_1,\zeta_1]$, if and only if
\begin{eqnarray}
\text{for $\Lambda =0$:} &&  k \,\geq \, 0\quad \text{and} \quad  c\, > \,0\;,
\label{first_condition1}
\\
\text{for $\Lambda >0$:} &&   k \,>\,  0\quad \text{and} \quad  c \,\in\, \mathbb{R}\;, \quad \text{or}
\nonumber
\\
&&    k \,= \, 0 \quad \text{and} \quad    c\, > \,0\;,
\label{first_condition2}
\\
\text{for $\Lambda <0$:} &&  -\frac{3}{\Lambda} \frac{c^2}{4} \,> \, k \,\geq   \, 0 \quad \text{and} \quad    c\, > \,0\;.
\label{first_condition3}
\end{eqnarray}
We observe that \eq{first_condition3} reduces to \eq{first_condition1}  as $\Lambda \nearrow 0$ whereas  \eq{first_condition2} does not as $\Lambda \searrow 0$.

Moreover, we then have (because of \eq{first_condition1}-\eq{first_condition3}  all square roots are real)
\begin{eqnarray}
\text{for $\Lambda =0$:} &&  \zeta_1 \,=\, \sqrt{ \frac{k}{c}}\;,
\label{first_condition1b}
\\
\text{for $\Lambda >0$:} &&  \zeta_1 \,=\, \sqrt{-\frac{3}{\Lambda}\frac{c}{2} + \sqrt{\Big( \frac{3}{\Lambda} \frac{c}{2}\Big)^2  + \frac{3}{\Lambda} k}}
\label{first_condition2b}
\;,
\\
\text{for $\Lambda <0$:} &&  \zeta_1 \,=\, \sqrt{-\frac{3}{\Lambda}\frac{c}{2} - \sqrt{\Big( \frac{3}{\Lambda} \frac{c}{2}\Big)^2  + \frac{3}{\Lambda} k}}
\label{first_condition3b}
\;.
\end{eqnarray}

Let us also take \eq{nablaZ_cond}, which holds automatically in the current setting, and Remark~\ref{remark_nablaZ_cond} into account. Then \eq{first_condition1}-\eq{first_condition3} become
\begin{eqnarray}
\text{for $\Lambda =0$:} && c\, > \,0
\quad  (\Longrightarrow \quad k  \,\geq\, 0)
\;,
\label{first_condition1b}
\\
\text{for $\Lambda >0$:} &&   c \,>\,  0 \quad  (\Longrightarrow \quad k\,\geq \, 0) \quad \text{or}
\nonumber
\\
&&      c\, \leq  \,0  \quad \text{and} \quad  k\,> \, 0\;,
\label{first_condition2b}
\\
\text{for $\Lambda <0$:} &&  c\, > \,0   \quad \text{and} \quad  -\frac{3}{\Lambda} \frac{c^2}{4} \,> \, k \,\geq   \, 0  \;.
\label{first_condition3b}
\end{eqnarray}
Finally, we need to make sure that   the image of the function $Z$ is  in $[-\zeta_1,\zeta_1]$:
\begin{enumerate}
\item[(i)]
For $\Lambda=0$ this will be the case if and only if $c Z^2 \leq k$, which follows automatically from  \eq{equation_k} and  \eq{nablaZ_cond}.
\item[(ii)]
 Due to \eq{nablaZ_cond} we have  $V(Z) = k -cZ^2 - \frac{\Lambda}{3}Z^4 \geq 0$, which, for $\Lambda >0$, implies that $Z^2$ is bounded by $\zeta_1^2$.
Indeed, using   \eq{first_condition2b}, $\pm \zeta_1$ are the only zeros of $V$, and $V$ is positive between them and  negative outside.
\item[(iii)]
For $\Lambda<0$ it follows from \eq{first_condition3b} that  $V$ has, in addition to $\pm \zeta_1$, two additional zeros $\pm \zeta_2$, with $\zeta_2>\zeta_1$.
$V$ is positive between $\pm \zeta_1$, and for  $\zeta>\zeta_2 $ and $\zeta<-\zeta_2$.
So either $Z^2\leq \zeta_1^2$, as desired, or
\begin{equation}
Z^2 \,\geq\, \zeta_2^2 \,=\,  -\frac{3}{\Lambda}\frac{c}{2} + \sqrt{\Big( \frac{3}{\Lambda} \frac{c}{2}\Big)^2  + \frac{3}{\Lambda} k}
\;.
\end{equation}
Taking \eq{nablaZ_cond} into account,  the latter inequality is equivalent to
\begin{equation}
 Z^2 + \frac{3}{\Lambda}\frac{c}{2} \,\geq\,0
\;.
\end{equation}
We conclude that
\begin{equation}
k \,\geq\,  cZ^2 + \frac{\Lambda}{3} Z^4 \, \geq \,  - \frac{3}{\Lambda}\frac{c^2}{4}
\;,
\end{equation}
 in contradiction to \eq{first_condition3b}. Thus
\begin{equation}
Z^2 \,\leq\, \zeta_1^2 \,=\,  -\frac{3}{\Lambda}\frac{c}{2} - \sqrt{\Big( \frac{3}{\Lambda} \frac{c}{2}\Big)^2  + \frac{3}{\Lambda} k}
\end{equation}
%
holds automatically. Equivalently (using  \eq{nablaZ_cond}),
%
\begin{equation}
 \frac{\Lambda}{3} Z^2+\frac{c}{2} \,\geq\,  0
\;,
\end{equation}
from which we deduce (using $c>0$) that $k\geq 0$, which we needed to assume in \eq{first_condition3b},
 holds in fact automatically.
\end{enumerate}
%
Altogether, we are led to the following corollary, which is a reformulation of a special case of Theorem~\ref{thm_charact_aux}
where the  NUT-parameter vanishes:
\begin{corollary}
\label{cor_charact_aux}
Let $(\mcM,g)$ be a smooth $(3+1)$-dimensional $\Lambda$-vacuum space-time which admits a KVF $X$ such that  the associated MST vanishes for some function $Q$.
Assume
 that $Q\mathcal{F}^2$ and $Q\mathcal{F}^2-4\Lambda$ are not identically zero,  and that $\mathrm{Im}\Big(\frac{\sqrt{\mathcal{F}^2}}{Q\mathcal{F}^2-4\Lambda}\Big)$ has non-zero gradient somewhere.
Assume further that    $b_2=0$ and that
\begin{eqnarray}
\text{for $\Lambda =0$:} && c\, > \,0
\quad  (\Longrightarrow \quad k  \,\geq\, 0)
\;,
\label{second_condition1}
\\
\text{for $\Lambda >0$:} &&   c \,>\,  0 \quad  (\Longrightarrow \quad k\,\geq \, 0) \quad \text{or}
\nonumber
\\
&&      c\, \leq  \,0  \quad \text{and} \quad  k\,> \, 0\;,
\label{second_condition2}
\\
\text{for $\Lambda <0$:} &&  c\, > \,0   \quad \text{and} \quad k\,<\, \frac{3}{|\Lambda |} \frac{c^2}{4}
\quad  (\Longrightarrow \quad k\,\geq \, 0)
\;.
\label{second_condition3}
\end{eqnarray}
Then $(\mcM, g)$ is locally isometric to a Kerr-(A)dS space-time with parameters $(\Lambda,m,a)$, where
\begin{equation}
m\,=\,\frac{b_1}{2\big(\frac{\Lambda}{3} \zeta_1^2 + c\big)^{3/2}}\,, \quad a\,=\, \frac{\zeta_1}{\big(\frac{\Lambda}{3} \zeta_1^2 + c\big)^{1/2}}\,.
\end{equation}
%
(Recall Proposition~\ref{constancy} which asserts that $c$ and $k$ are constants.)
The Schwarzschild-(A)dS limit is obtained for $a=0$, equivalently $k=0$.
\end{corollary}

\begin{remark}
{\rm
For $\Lambda=0$, Corollary~\ref{cor_charact_aux} reduces to Theorem~\ref{thm_charact1} (ii), albeit
 Corollary~\ref{cor_charact_aux} makes, in addition, concrete predictions concerning the parameters of the  Kerr space-time one is dealing with.
The assumption $l>0$ made there corresponds to $b_2=0$, while $-|X|^2-\mathrm{Re}(\chi)>0$ corresponds to $c>0$.

For $\Lambda>0$  Corollary~\ref{cor_charact_aux}  is in accordance with  the results  in \cite{mpss2}: For $b_2=0$ and $c>0$ there is just the KdS space-time (including SdS), while for $b_2=0$ and  $c\leq 0 $ there are other $\Lambda>0$-vacuum space-times with vanishing MST such as the Kottler space-times with $\varepsilon$ equals $0$ or $-1$, or the Wick-rotated Kerr-AdS space-time, whence the additional condition $k>0$ is indeed necessary.
}
\end{remark}

\section{The function $Q$}
\label{sec_function_Q}

As a matter of fact, in order to calculate the MST one needs to fix the function $Q$.
As already indicated in the Introduction and as  became evident  in Section~\ref{sec_results},
space-times of particular interest are those for which the MST vanishes.
The most natural way to choose $Q$ is to require a certain component of the MST to vanish (or one of its derivatives), and to solve the corresponding equation for $Q$,
which usually requires the non-vanishing of certain scalars.

Proceeding this way one is led to a variety  of  $Q$'s (cf.\ below), depending on the metric $g$, the KVF $X$ and derivatives thereof, which come along with
different useful properties.
In general, all these different choices are non-equivalent.
Now, it immediately follows from \eq{dfn_mars-simon} that if there exist two functions $Q$ which differ on an open set such that the MST vanishes, then the space-time is Weyl-flat on this set. Excluding this possibility, we see that 
in the setting of a vanishing MST  all  choices of $Q$ are equivalent,
and thus provide very useful relations
between $g$, $X$ and fields composed of them,
which \emph{necessarily} need to be fulfilled in vacuum space-times with vanishing MST (or on initial data sets which are supposed to generate such space-times \cite{mpss, kerr_part2}).
These relations  will be key for the algorithmic approach described in Section~\ref{sec_complex_KVFs}.

\subsection{``Classical'' definitions of $Q$ and some properties}
\label{sec_def_Q}

We consider a smooth $(3+1)$-dimensional $\Lambda$-vacuum space-time $(\mcM,g)$ which admits a KVF $X$.
Two different definitions of $Q$ have been beneficial in the literature. They have decisive advantages depending on the context where the MST is used (cf.\ \cite{mpss}).
The first and most elementary
 definition of $Q$ is obtained if we require the scalar $\mathcal{S}(\mathcal{F},\mathcal{F})$ to vanish,
\begin{equation}
0 = \mathcal{F}^{\mu\nu}\mathcal{F}^{\sigma\rho} \mathcal{S}_{\mu\nu\sigma\rho}
= \mathcal{F}^{\mu\nu}\mathcal{F}^{\sigma\rho}( \mathcal{C}_{\mu\nu\sigma\rho}  + Q\mathcal{Q}_{\mu\nu\sigma\rho})
=\mathcal{F}^{\mu\nu}\mathcal{F}^{\sigma\rho}\mathcal{C}_{\mu\nu\sigma\rho}  -\frac{2}{3} Q\mathcal{F}^4
\,.
\end{equation}
Restricting attention to those regions of space-time where
\begin{equation}
\mathcal{F}^2\, \ne\, 0
\label{non_zero}
\end{equation}
%
one is immediately led to the following definition of $Q$:
\begin{equation}
Q_0  \,:=\,  \frac{3}{2}\mathcal{F}^{-4} \mathcal{F}^{\mu\nu}\mathcal{F}^{\sigma\rho}\mathcal{C}_{\mu\nu\sigma\rho}
\,.
\label{1st_dfn_Q}
\end{equation}
Whenever there exists a function $Q$ such that the MST vanishes,
on the set on which \eq{non_zero} holds
$Q$ has to be given by \eq{1st_dfn_Q}.

Let us  devote attention to a second significant definition of $Q$.
For the time being, let us think of a Cauchy problem.
The choice $Q=Q_0$ allows us to derive necessary conditions on initial data sets to end up with a vacuum space-time with vanishing MST.
To derive conditions which are \emph{sufficient} as well, one needs evolution equations for the MST (more precisely, one needs a homogeneous  set of equations which ensures that, given an appropriate set of zero initial data, the zero-solution is the only one). So far there are no such equations available
for an MST satisfying $Q=Q_0$.

We therefore need an alternative definition of the function $Q$ which allows the derivation of a symmetric hyperbolic system
of evolution equations for the MST in a vacuum space-time.
This was accomplished in \cite{IK} for $\Lambda=0$.
A generalization  which allows the derivation of evolution equations for any sign of the cosmological constant was provided in \cite{mpss}.
Albeit the evolution equations will be of  importance  only in part II \cite{kerr_part2},  it is useful to have both ``classical'' definitions of  $Q$ at disposal:
Firstly   because we want to analyze the relation between the various definitions of $Q$, and secondly because
 it turns out that the alternative definition of $Q$, and in  particular the PDE which this choice of $Q$ satisfies, will be   very  useful later on.

Let us assume
that $(\mcM,g)$  admits a KVF $X$ which satisfies  \eq{non_zero}
and for which  the associated  MST vanishes for some function~$Q$,
\begin{equation}
\mathcal{S}_{\alpha\beta\mu\nu} \,\equiv\, \mathcal{C}_{\alpha\beta\mu\nu} - Q (\mathcal{F}_{\alpha\beta}\mathcal{F}_{\mu\nu}
- \frac{1}{3}\mathcal{F}^2\mathcal{I}_{\alpha\beta\mu\nu}) \,=\, 0
\,.
\label{vanishing_S}
\end{equation}
%
For the following computations recall the formulas in \eq{self_dual}-\eq{useful_last}.
Contraction of \eq{vanishing_S} with $\mathcal{F}^{\alpha\beta}$ yields
\begin{equation}
\mathcal{F}^{\alpha\beta}\mathcal{C}_{\alpha\beta\mu\nu} \,=\, \frac{2}{3} Q \mathcal{F}^2\mathcal{F}_{\mu\nu}
\,.
\label{contr_vanishing_S}
\end{equation}
We  use  \eq{vanishing_S}-\eq{contr_vanishing_S} together with \eq{ernst_form} to rewrite
\eq{rln_nablaF} and \eq{rln_nablaF2} (cf.\ \cite{mars_senovilla}),
\begin{eqnarray}
\nabla_{\mu}\mathcal{F}_{\alpha\beta} &=& \frac{1}{2}Q\chi_{\mu} \mathcal{F}_{\alpha\beta}
+\frac{1}{3} (Q\mathcal{F}^2 - 4\Lambda )X^{\nu}\mathcal{I}_{\alpha\beta\mu\nu}
\,,
\label{rln_nablaF_2}
\\
\nabla_{\mu}\mathcal{F}^2 &=&  \frac{2}{3} (Q \mathcal{F}^2  +2\Lambda) \chi_{\mu}
\,.
\label{rln_nablaF2_2}
\end{eqnarray}
Let us compute the divergence of \eq{vanishing_S},
\begin{eqnarray}
0&=&\nabla^{\nu}Q (\mathcal{F}_{\alpha\beta}\mathcal{F}_{\mu\nu}
- \frac{1}{3}\mathcal{F}^2\mathcal{I}_{\alpha\beta\mu\nu})
+  Q \mathcal{F}_{\mu}{}^{\nu}\nabla_{\nu}\mathcal{F}_{\alpha\beta}
+ Q\mathcal{F}_{\alpha\beta}\nabla_{\nu}\mathcal{F}_{\mu}{}^{\nu}
\nonumber
\\
&& - \frac{1}{3}Q\nabla_{\nu}\mathcal{F}^2\mathcal{I}_{\alpha\beta\mu}{}^{\nu}
\\
&=&\nabla^{\nu}Q (\mathcal{F}_{\alpha\beta}\mathcal{F}_{\mu\nu}
- \frac{1}{3}\mathcal{F}^2\mathcal{I}_{\alpha\beta\mu\nu})
- \frac{1}{3}(Q\mathcal{F}^2 -  4\Lambda ) Q \mathcal{F}_{\mu}{}^{\nu} X^{\gamma}\mathcal{I}_{\alpha\beta\gamma\nu}
\nonumber
\\
&&
+\frac{1}{4}Q( \mathcal{F}^2+ 4\Lambda )  X_{\mu} \mathcal{F}_{\alpha\beta}
- \frac{2}{9}Q(
Q\mathcal{F}^2+2\Lambda )\chi_{\nu} \mathcal{I}_{\alpha\beta\mu}{}^{\nu}
\,.
\end{eqnarray}
Contracting this with $\mathcal{F}^{\alpha\beta}$ and using  \eq{non_zero} we are led to the equation
\begin{equation}
\mathcal{F}_{\mu}{}^{\nu}\nabla_{\nu}Q
+ \frac{1}{12}(Q\mathcal{F}^2   +20\Lambda ) Q X_{\mu} \,=\, 0
\,.
\end{equation}
Contraction with $\mathcal{F}^{\mu}{}_{\kappa}$
yields
\begin{equation}
\mathcal{F}^2\nabla_{\kappa}Q
+ \frac{1}{6}(Q\mathcal{F}^2   +20\Lambda ) Q\chi_{\kappa}\,=\, 0
\,.
\label{PDE_Q}
\end{equation}
Assuming
\begin{equation}
Q\mathcal{F}^2\ne 0\;, \quad Q\mathcal{F}^2-4\Lambda \,\ne\, 0
\,.
\label{non_zero2b}
\end{equation}
it follows from \eq{rln_nablaF2_2} that \eq{PDE_Q} is equivalent to
%
%
\begin{equation}
 \frac{Q\mathcal{F}^2+2\Lambda}{Q\mathcal{F}^2-4\Lambda } \nabla_{\kappa}\log (Q\mathcal{F}^2)   -\frac{3}{4}\nabla_{\kappa}\log\mathcal{F}^2 \,=\, 0
\;,
\label{PDE_Q2b}
\end{equation}
Let us consider the $\Lambda=0$-case first. Then \eq{PDE_Q2b}
becomes
%
\begin{equation}
  \nabla_{\kappa}Q
 \,=\, -  \frac{1}{4} Q\mathcal{F}^{-2 }\nabla_{\kappa} \mathcal{F}^2
\label{PDE_Lambda0}
\end{equation}
The general solution of this equation reads
%
\begin{equation}
Q_{\widehat\varsigma }\,=\, \widehat\varsigma (-\mathcal{F}^2)^{-1/4}
 \,, \quad \widehat\varsigma\in\mathbb{C}
\label{anotherQ}
\end{equation}
(compare \cite{IK2}, where such a choice of $Q$ appears), and provides a possible choice for $Q$ in the $\Lambda=0$-case.

For an arbitrary cosmological constant we observe that, using  \eq{rln_nablaF2_2},  the differential  equation \eq{PDE_Q} is equivalent to
\begin{equation}
\nabla_{\kappa}(Q\mathcal{F}^2)
- \frac{1}{2}(Q\mathcal{F}^2   -4\Lambda   ) Q\nabla_{\kappa}\chi\,=\, 0
\,.
\label{gen_PDE_Q}
\end{equation}
Still assuming \eq{non_zero2b}, we set
%
\begin{equation}
\mathcal{A} \,:=\, 6\mathcal{F}^{2} \frac{Q\mathcal{F}^2 +2\Lambda}{(Q\mathcal{F}^2-4\Lambda)^2}
\,.
\label{dfn_A}
\end{equation}
Using  \eq{rln_nablaF2_2} and \eq{gen_PDE_Q}, one readily checks that
\begin{equation}
\nabla_{\kappa}(\mathcal{A}-\chi)\,=\,0
\;, \quad
\Longleftrightarrow
\quad
\mathcal{A} \,=\, \chi  + \lambda \,, \quad \lambda \in\mathbb{C}
\,.
\label{relation_A_chi}
\end{equation}
Recall  that the Ernst potential $\chi$ is  only defined  up to some additive complex constant, the $\chi$-constant, which we have not specified yet.
Due to this freedom we may assume without restriction that the integration constant $\lambda$ vanishes.
Once we know $\mathcal{A}$, the function $Q$ can be determined from  \eq{dfn_A}  by solving a quadratic equation,
\begin{equation}
Q^2- 2Q\mathcal{F}^{-2}(4\Lambda +3\mathcal{F}^2 \chi^{-1}) +4\Lambda \mathcal{F}^{-4}(4\Lambda    - 3\mathcal{F}^{2}\chi^{-1})\,=\,0
\label{PDE_Q_Lambda}
\;.
\end{equation}
The solution provides a second important choice for $Q$,
\begin{eqnarray}
Q_{\mathrm{ev}}\,:=\,\frac{3\mathcal{F}^2  + 4\Lambda \chi \pm
 3\sqrt{\mathcal{F}^2(\mathcal{F}^2 + 4\Lambda \chi)}}{\chi\mathcal{F}^{2}}
\,,
\label{anotherQLambda}
\label{evQ_with_Lambda}
\end{eqnarray}
supposing that
\begin{equation}
\chi\,\ne\, 0 \quad  \overset{\eq{non_zero2b}}{\Longleftrightarrow}  \quad  Q\mathcal{F}^2 +2\Lambda \ne 0
\,.
\label{non_zero3}
\end{equation}
In the case of a vanishing cosmological constant $\Lambda$, we set, away from the zero-set of $\chi$,
\begin{equation}
Q_{\mathrm{ev}} \,:=\,  \frac{6}{\chi}
\,,
\label{evQ_without_Lambda}
\end{equation}
which is a special case of \eq{evQ_with_Lambda}.

A remark is in order concerning the appearance of  roots in \eq{evQ_with_Lambda} and many other equations below:
The assumptions \eq{non_zero} and \eq{add_ass_Q} below imply that the term under the square root does not have any zeros.
Since  therefore no branch point is ever met, one may prescribe a square root at one point and extend it by continuity to the whole manifold.
That yields  smooth functions $\sqrt{\,\cdot\,}$.
\label{discussion_root}

In this work we are exclusively interested in space-times with vanishing MST.
The derivation of  $Q_{\mathrm{ev}}$ reveals  that  in  such space-times
there exists an
Ernst potential and a choice of the sign in \eq{anotherQLambda}  such that  $Q(=Q_0)$ equals $Q_{\mathrm{ev}}$, supposing that
 \eq{non_zero2b} and \eq{non_zero3} hold
 (compare \cite{mpss}, where, for $\Lambda >0$, conditions  are derived which \emph{characterize}
the setting where the leading order terms of $Q_0$ and $Q_{\mathrm{ev}}$ coincide on $\scri$).
The particular  choice of the  sign in \eq{anotherQLambda} crucially depends  on the asymptotic behavior of the corresponding functions, and thereby on the sign
of the cosmological constant \cite{mpss}.
So, although many subsequent formulas involve a ``sign ambiguity'' $\pm$, in the setting of a vanishing MST it is always only one solution which is
relevant after all.


Taking Proposition~\ref{constancy} and Theorem~\ref{thm_charact1} into account we are led to the following result (cf.\ \cite[Theorem 4]{mars_senovilla}), which shows that
as long as we focus attention to space-times with vanishing MST we can use $Q_0$, $Q_{\mathrm{ev}}$ (and $Q_{\widehat\varsigma }$ in the $\Lambda=0$ case) interchangeably.

%
\begin{proposition}
\label{prop_rln_Q}
Consider  a smooth $(3+1)$-dimensional $\Lambda$-vacuum space-time which admits a (possibly complex) KVF.
\begin{enumerate}
\item[(i)]
Assume that the MST vanishes for some function $Q$, that  $Q\mathcal{F}^2+2\Lambda \ne 0$  everywhere,  that
 $Q\mathcal{F}^2$ and  $Q\mathcal{F}^2-4\Lambda$ are not identically zero, and that $\mathrm{Im}\Big(\frac{\sqrt{\mathcal{F}^2}}{Q\mathcal{F}^2-4\Lambda}\Big)$ has non-zero gradient somewhere.
Then
there exists an Ernst potential $\chi$ and a choice of $\pm$ in \eq{anotherQLambda}  such that
\begin{equation}
Q\,=\, Q_0 \,=\, Q_{\mathrm{ev}}
\,.
\end{equation}
In particular, the function $Q$ satisfies  the PDE \eq{PDE_Q2b}.

\item[(ii)]
Assume that the MST vanishes for some function $Q$, that $\Lambda=0$, and that $\mathcal{F}^2$ is not identically zero.
Then
there exists an $\widehat\varsigma \in\mathbb{C}$ such that
\begin{equation}
Q\,=\, Q_0 \,=\, Q_{\widehat\varsigma }
\,.
\end{equation}
\end{enumerate}
\end{proposition}

\begin{remark}
{\rm
In Section~\ref{sec_yet_another} we will introduce further  choices of the function $Q$ which likewise will be shown to coincide altogether in the setting of a vanishing MST.
}
\end{remark}

\begin{remark}
{\rm
When dealing with the MST one usually assumes  the existence of a \emph{real} KVF.
At this stage it does not matter whether $X$ is real or complex.
Of course, unless its real and imaginary part are linearly dependent, the existence of a complex KVF implies the existence of two real KVFs.
Note, however, that the existence of a real KVF w.r.t.\ which the MST vanishes, also implies the existence of a second real KVF \cite{mars_senovilla}.
The fact that $X$ might  be complex does not cause any problems. In fact,
all the relations derived so far remain true for complex Killing vector fields without any modification, in particular the definition of the MST itself
is meaningful.
}
\end{remark}

As already indicated above, the crucial property of the choice $Q=Q_{\mathrm{ev}}$ is  that it comes along with hyperbolic  evolution equations for the MST.
Assuming that, in addition to \eq{non_zero} and \eq{non_zero3},
\begin{equation}
Q_{\mathrm{ev}}\mathcal{F}^2 + 8\Lambda \,\ne \, 0
\;,
\label{add_ass_Q}
\end{equation}
(which implies $\mathcal{F}^2 + 4\Lambda \chi \ne 0$),
the associated MST satisfies a regular, linear, homogeneous  symmetric hyperbolic system \cite{mpss}
\begin{eqnarray}
\nabla_{\beta} \mathcal{S}^{(\mathrm{ev})}_{\mu\nu\alpha}{}^{\beta}
&=&
 - Q_{\mathrm{ev}}\Big( \mathcal{F}_{\alpha\beta}\delta_{\mu}{}^{\gamma}\delta_{\nu}{}^{\delta}   -  \frac{2}{3}
  \mathcal{F}^{\gamma\delta} \mathcal{I}_{\alpha\beta\mu\nu}
\Big)X^{\lambda}\mathcal{S}^{(\mathrm{ev})}_{\gamma\delta\lambda}{}^{\beta}
\nonumber
\\
&&
\qquad
-4 \Lambda  \frac{  5  Q_{\mathrm{ev}}\mathcal{F}^2  +4\Lambda }{Q_{\mathrm{ev}}\mathcal{F}^2 + 8\Lambda}
 \mathcal{Q}_{\mu\nu\alpha\beta}
   \mathcal{F}^{-4}\mathcal{F}^{\gamma\delta} X^{\lambda}   \mathcal{S}^{(\mathrm{ev})}_{\gamma\delta\lambda}{}^{\beta}
\;.
\label{phys_ev}
\end{eqnarray}

From  equation \eq{phys_ev} one  further derives straightforwardly  a linear, homogeneous system of wave equations of the form
\begin{equation}
\Box_g \mathcal{S}_{\mu\nu\sigma\rho} = \mathcal{M}(\mathcal{S},\nabla\mathcal{S})_{\mu\nu\sigma\rho}
\,,
\label{scem_wave}
\end{equation}
which is satisfied by the MST \cite{IK, mpss}.
With regard to the derivation of evolution equations
the ambiguity in the definition of $Q_{\mathrm{ev}}$, which arises from  the freedom to choose the $\chi$-constant and the
sign ambiguity in \eq{anotherQLambda},
does not matter.

For vanishing cosmological constant
 \eq{phys_ev} simplifies to
\begin{eqnarray}
\nabla_{\beta}\mathcal{S}^{(\mathrm{ev})}_{\mu\nu\alpha}{}^{\beta}
\,=\, -Q_{\mathrm{ev}} \Big(\mathcal{F}_{\alpha\beta}\delta_{\mu}{}^{\gamma}\delta_{\nu}{}^{\delta} - \frac{2}{3} \mathcal{F}^{\gamma\delta}\mathcal{I}_{\alpha\beta\mu\nu}\Big)
 X^{\lambda}\mathcal{S}^{(\mathrm{ev})}_{\gamma\delta\lambda}{}^{\beta}
\,.
\label{shsee}
\end{eqnarray}
 This system  has already been derived in \cite{IK}. Note that in this setting \eq{non_zero2b} and  \eq{add_ass_Q}  can simply replaced by \eq{non_zero}.

A disadvantage of $Q_{\mathrm{ev}}$ is that it generally cannot be computed explicitly since
it involves the Ernst potential whose computation requires an integration.
A decisive exception is provided by the $\Lambda=0$-case, where this can be done in terms of $\mathcal{F}^2=\mathcal{F}^2(g, \nabla X)$, supposing that attention is restricted to space-times with vanishing MST, cf.\ Theorem~\ref{thm_charact1}.%
\footnote{
In fact, the Ernst potential is not needed at all, since \eq{anotherQ} can be employed to express $Q$ directly  in terms of $\mathcal{F}^2$.
For $\Lambda\ne 0 $ a corresponding result does not seem to exist.
}

\subsection{Yet two alternative definitions of   $Q$}
\label{sec_yet_another}

It turns out that   in view of the algorithmic
approach we are aiming at, the choices $Q_0$ and $Q_{\mathrm{ev}}$ for $Q$ are not convenient.
For this reason there remains the need to look for  alternative  choices  which are more suitable
for our purposes.

Our ultimate aim is to solve \eq{vanishing_S} for $X$, i.e.\ to derive an (algebraic) expression for $X$ in terms of $g$ (and derivatives thereof) which necessarily needs to hold
in any  space-time  which admits a KVF w.r.t.\ which the MST vanishes.

In Section~\ref{sec_def_Q} we have given a definition of the function $Q$, \eq{1st_dfn_Q}, which we denoted by $Q_0$, and which naturally arises if a certain component of the MST is  required to vanish.
In this section we provide a related  definition of the function $Q$, \eq{yet_another_Q} below.
If we further  exploit the fact that, in the case of a vanishing MST, $Q$ satisfies the PDE \eq{PDE_Q2b},
 we are led to a very convenient definition of $Q$, \eq{yet_another_Q_2}.
It will have the decisive advantage that  it can be given explicitly just in terms
of the self-dual Weyl tensor,
 and does not rely on the KVF.

A vanishing MST $\mathcal{S}_{\alpha\beta\mu\nu}$ implies by contraction with $\mathcal{F}^{\alpha\beta}\mathcal{F}^{\mu\nu}$,
\begin{equation}
Q\mathcal{F}^4 \,=\,  \frac{3}{2} \mathcal{F}^{\alpha\beta} \mathcal{F}^{\mu\nu}\mathcal{C}_{\alpha\beta\mu\nu}
\;,
\label{intermediate_rln}
\end{equation}
while contraction with $\mathcal{C}^{\alpha\beta\mu\nu}$ yields, due to the algebraic properties of the Weyl tensor,
\begin{equation}
 Q \mathcal{F}^{\alpha\beta}\mathcal{F}^{\mu\nu} \mathcal{C}_{\alpha\beta\mu\nu}
  \,= \, \mathcal{C}^2
\;,
\end{equation}
where we have set
\begin{equation}
\mathcal{C}^2\,:=\,   \mathcal{C}^{\alpha\beta\mu\nu}\mathcal{C}_{\alpha\beta\mu\nu}
\;.
\end{equation}
Combined, we obtain
\begin{equation}
Q^2 \mathcal{F}^4 \,=\,  \frac{3}{2}  \mathcal{C}^2
\;.
\label{rln_Q2F4_C2}
\end{equation}
For $\mathcal{F}^2\ne 0$ this provides another possible definition of the function $Q$,
%
\begin{equation}
Q_{\mathcal{F}} \,:=\,  \pm
\sqrt{ \frac{3}{2}}\,\mathcal{F}^{-2} \sqrt{\mathcal{C}^2}
\;.
\label{yet_another_Q}
\end{equation}
Such a sign  ambiguity already appeared in the definition of $Q_{\mathrm{ev}}$. It disappears
if the MST is required to vanish:
Similar to  $Q_{\mathrm{ev}}$, it follows from its derivation  that  the function $Q$  necessarily coincides with $Q_{\mathcal{F}}$, either with $+$ or $-$
in \eq{yet_another_Q}.

Next, we work on the set where
\begin{equation}
Q\mathcal{F}^2\,\ne 0\;, \quad Q\mathcal{F}^2-4\Lambda \ne 0\;,
\label{int_assumptions_Q}
\end{equation}
and observe that \eq{PDE_Q2b} is equivalent to
\begin{equation}
\frac{3}{2}\nabla_{\kappa}\log\mathcal{F}^2 \,=\,  \nabla_{\kappa}\log\bigg[
\frac{(Q\mathcal{F}^2 - 4\Lambda)^3}{Q\mathcal{F}^2}
\bigg]
\;.
\end{equation}
This differential equation can be integrated,
\begin{equation}
\mathcal{F}^2 \,=\,  \widehat\varkappa
\frac{(Q\mathcal{F}^2 - 4\Lambda)^2}{(Q\mathcal{F}^2)^{2/3}} \;, \quad  \widehat\varkappa\in\mathbb{C}\setminus\{0\}
\;.
\label{solution_DE_Q}
\end{equation}
In particular, $Q=Q_{\mathcal{F}}$ needs to satisfy this equation in space-times with vanishing MST, so let us plug it in,
%
\begin{equation}
 \mathcal{F}^2
\,=\, \widehat\varkappa\Big(\frac{3}{2}\Big)^{2/3} (\mathcal{C}^2)^{-1/3} \Big( \pm \sqrt{\mathcal{C}^2} -
\sqrt{\frac{32}{3}}\,\Lambda\Big)^2 \;, \quad  \widehat\varkappa\in\mathbb{C}\setminus\{0\}
\,.
\label{rln_C2_F2_2}
\end{equation}
In terms of $\mathcal{C}^2$ the assumptions \eq{int_assumptions_Q} become
\begin{eqnarray}
 \mathcal{C}^2 \,\ne \,0
\;, \quad
\mathcal{C}^2  -
\frac{32}{3}\Lambda^2 \,\ne \, 0
\;.
\label{ineqs_C2_0}
\end{eqnarray}
The latter condition can be replaced by $\pm \sqrt{\mathcal{C}^2}   -
 \sqrt{\frac{32}{3}}\Lambda \ne  0$, and merely needs to hold for one sign, depending on the sign in \eq{yet_another_Q_2} below  for which the MST vanishes.

From \eq{yet_another_Q} and \eq{rln_C2_F2_2} one further obtains  
\begin{equation}
 Q_{\mathcal{C}} \,:=\, \varkappa (\mathcal{C}^2)^{5/6}\Big( \pm\sqrt{\mathcal{C}^2} -
\sqrt{\frac{32}{3}}\,\Lambda\Big)^{-2} \;, \quad  \varkappa\in\mathbb{C}\setminus\{0\}
\;,
\label{yet_another_Q_2}
\end{equation}
which provides yet another definition of the function $Q$ which necessarily needs to be fulfilled if the corresponding MST vanishes (again, for one sign $\pm$). Recall the prescription of roots on page~\pageref{discussion_root}.
As for $Q_{\mathrm{ev}}$, defined in Section~\ref{sec_def_Q}, this is actually a 1-parameter family and the MST, \emph{for a given KVF $X$}, vanishes for at most one of them.
A decisive advantage of this definition of $Q$ is that  it does \emph{not} depend on the KVF,
apart from the scale factor $\varkappa$ which is related to the length of the KVF, cf.\ below.

In terms of $\mathcal{C}^2$ the conditions  $Q\mathcal{F}^2+2\Lambda\ne0 $ and $Q\mathcal{F}^2+8\Lambda\ne0 $
read
\begin{eqnarray}
\mathcal{C}^2   -\frac{8}{3}\,\Lambda^2 \,\ne \, 0
\;, \quad
\mathcal{C}^2 - \frac{128}{3}\,\Lambda^2 \,\ne \, 0
%
\;.
\label{ineqs_C2_1}
\end{eqnarray}
As above, these conditions  may be replaced by $\pm\sqrt{\mathcal{C}^2}   + \sqrt{\frac{8}{3}}\,\Lambda \ne 0$ and
$\pm\sqrt{\mathcal{C}^2}   + \sqrt{\frac{128}{3}}\,\Lambda \ne  0$, respectively.
As an extension of Proposition~\ref{prop_rln_Q} we  obtain:
\begin{proposition}
\label{prop_rln_Q2}
Assume that the MST vanishes for some function $Q$,
and that  the inequalities \eq{ineqs_C2_0} hold.
Then there exists
a constant $\varkappa \in\mathbb{C}\setminus\{0\}$ and a
choice of $\pm$ in the corresponding expressions
such that
\begin{equation}
Q\,=\, Q_0 \,=\, Q_{\mathcal{F}} \,=\,  Q_{\mathcal{C}}
\,.
\end{equation}
Assume that, in addition, \eq{ineqs_C2_1} holds.
Then
there exists an Ernst potential $\chi$ for $ Q_{\mathrm{ev}}$ such that
\begin{equation}
Q\,=\, Q_0 \,=\, Q_{\mathrm{ev}}\,=\, Q_{\mathcal{F}} \,=\,  Q_{\mathcal{C}}
\,.
\end{equation}
\end{proposition}

A remark concerning the role of the constant $\varkappa$ is in order:
The MST vanishes w.r.t.\ the KVF $X$ if and only if it vanishes w.r.t.\ the KVF $\lambda X$, $\lambda\in\mathbb{C}\setminus \{0\}$.
Then $Q_0\mapsto \lambda^{-2} Q_0$ and $\varkappa\mapsto \lambda^{-2}\varkappa$.
The constant $\varkappa$ thus represents a ``scale factor''. It   may be regarded as a gauge constant  which
reflects the freedom to rescale the KVF.
 It  can be set equal to 1
(some care is needed when attention is restricted to real KVFs; we will come back to this issue in Section~\ref{sec_van_real_KVF}).


\section{Complex KVFs with vanishing MST}
\label{sec_complex_KVFs}

 Assume we have been given a  smooth $(3+1)$-dimensional
$\Lambda$-vacuum space-time $(\mcM,g)$  and a KVF $X$
such that the associated MST vanishes for some function $Q$.
Assuming further   $\mathcal{F}^2\ne 0$ and $Q\mathcal{F}^2 +2\Lambda\ne 0$ (one may employ Proposition~\ref{constancy} to relax the first  assumption),
it follows from \eq{rln_nablaF2_2} and \eq{PDE_Q} that
 the function $Q$ satisfies the PDE
 \begin{equation}
\mathcal{F}^2\partial_{\kappa}Q
+ \frac{1}{4}\frac{Q\mathcal{F}^2   +20\Lambda }{Q\mathcal{F}^2 +2\Lambda} Q\partial_{\kappa}\mathcal{F}^2\,=\, 0
\,.
\label{PDE_Q2}
\end{equation}
We want to express  $X^{\mu}$ in terms of   $\mathcal{F}_{\mu\nu}$ and $g_{\mu\nu}$.
In fact, this can be done straightforwardly via
\eq{rln_chi} and \eq{rln_nablaF2_2},
\begin{equation}
 X_{\mu} \,=\, 2 \mathcal{F}^{-2} \chi_{\alpha}\mathcal{F}_{\mu}{}^{\alpha}
\\
\,=\, 3(Q \mathcal{F}^2  +2\Lambda)^{-1} \mathcal{F}^{-2}  \mathcal{F}_{\mu}{}^{\alpha} \nabla_{\alpha}\mathcal{F}^2
\,.
\label{Killing_cand}
\end{equation}
If a space-time admits a KVF $X$  such that its  associated MST vanishes and such that $\mathcal{F}^2\ne 0$ and $Q\mathcal{F}^2 +2\Lambda\ne 0$, it \emph{necessarily} needs to be of this form.

For the remainder of this section we do \emph{neither} assume that  $\mathcal{F}_{\mu\nu}$ arises from a KVF \emph{nor}
that the space-time admits a KVF.
Instead, we  assume that a $\Lambda$-vacuum space-time $(\mcM,g)$ admits a self-dual two-form  $\mathcal{F}_{\mu\nu}$
and a function $Q=Q(\Lambda, \mathcal{F}^2)$ such that the following relations hold:
 \begin{eqnarray}
\mathcal{S}_{\alpha\beta\mu\nu}  \,\equiv \, \mathcal{C}_{\alpha\beta\mu\nu} - Q (\mathcal{F}_{\alpha\beta}\mathcal{F}_{\mu\nu}
- \frac{1}{3}\mathcal{F}^2\mathcal{I}_{\alpha\beta\mu\nu}) \,=\, 0
\,,
\label{relation1}
\\
\mathcal{F}^2\partial_{\kappa}Q
+ \frac{1}{4}\frac{Q\mathcal{F}^2   +20\Lambda }{Q\mathcal{F}^2 +2\Lambda} Q\partial_{\kappa}\mathcal{F}^2\,=\, 0
\,,
\label{relation2}
\end{eqnarray}
where we assume that everywhere (in the region of interest)
%
\begin{equation}
\mathcal{F}^2\,\ne\,0\,,
 \quad Q \mathcal{F}^2   -4  \Lambda\ne 0\;,
\quad  Q\mathcal{F}^2 +2\Lambda\,\ne\,0
\;.
\label{relation3}
\end{equation}
It seems worth emphasizing  that \eq{relation2} is \emph{not} a consequence of \eq{relation1} anymore  since we do \emph{not} assume $\mathcal{F}_{\mu\nu}$  to arise from a KVF.

Our aim is to construct a vector field on $\mcM$ (or at most finitely many) as the only KVF candidate whose associated MST vanishes, and which can be expressed just in terms of the metric $g$
(and derivatives thereof).

\subsection{Petrov type D }
\label{sec_petrov}

To do that, we first want to solve the condition of a vanishing MST for the self-dual two-form $\mathcal{F}_{\mu\nu}$.
For this the results from Section~\ref{sec_yet_another} are crucial,
 which allow us to express the function $Q$, and also  $\mathcal{F}^2$, in terms of the Weyl tensor,
 without any knowledge about the KVF.

%
\begin{lemma}
\label{relation_Q_diffQ}
Let $(\mcM, g)$ be a smooth $(3+1)$-dimensional $\Lambda$-vacuum space-time which admits a self-dual two-form $\mathcal{F}_{\mu\nu}$
and a function $Q$ such that $\mathcal{S}_{\alpha\beta\mu\nu}=0$.
Then the following statements are equivalent:
\begin{enumerate}
\item[(i)] $Q$ satisfies the PDE \eq{relation2}, and the inequalities  \eq{relation3} are fulfilled.
\item[(ii)]  $\mathcal{C}^2$ and $\mathcal{F}^2$ are related via \eq{rln_C2_F2_2}, and the following  inequalities  are fulfilled
%
\begin{equation}
 \mathcal{C}^2 \,\ne \,0
\;, \quad
\pm \sqrt{\mathcal{C}^2}   -
 \sqrt{\frac{32}{3}}\Lambda \,\ne \, 0
\;, \quad
\pm \sqrt{\mathcal{C}^2}   +
 \sqrt{\frac{8}{3}}\Lambda \,\ne \, 0
\;.
\label{ineqs_C2_new}
\end{equation}
\item[(iii)] $Q=Q_{\mathcal{C}}$, and the inequalities \eq{ineqs_C2_new} are fulfilled.
\end{enumerate}
\end{lemma}
\begin{proof}
We observe that for the considerations in Section~\ref{sec_yet_another} it does not matter whether the self-dual two-form $\mathcal{F}_{\mu\nu}$ arises from a KVF.
It has been shown there that \eq{relation1} implies  $Q=Q_{\mathcal{F}}$.
Moreover, it follows from  Section~\ref{sec_yet_another} that \eq{relation2} is equivalent to  \eq{solution_DE_Q}, which, using $Q=Q_{\mathcal{F}}$,
is equivalent to  \eq{rln_C2_F2_2}, and thus also to $Q=Q_{\mathcal{C}}$.
\qed
\end{proof}



Assume that  \eq{ineqs_C2_new} holds (in fact for this part it suffices if \eq{ineqs_C2_0} holds).
 We consider \eq{relation1}, which we regard as an equation for
$\mathcal{F}_{\mu\nu}$,   and we take
\begin{equation}
 Q\,=\, Q_{\mathcal{C}} \,\equiv\,   \varkappa(\mathcal{C}^2)^{5/6}\Big( \pm\sqrt{\mathcal{C}^2}-
\sqrt{\frac{32}{3}}\,\Lambda\Big)^{-2}
\;,
\label{cand_Q}
\end{equation}
for some  choice of $\varkappa\in\mathbb{C}\setminus \{0\}$.
%
%
%
It follows from Lemma~\ref{relation_Q_diffQ} that
\begin{equation}
\mathcal{F}^2 \,=\, \pm
\varkappa^{-1}\sqrt{\frac{3}{2}}(\mathcal{C}^2)^{-1/3}\Big( \pm \sqrt{\mathcal{C}^2}-
\sqrt{\frac{32}{3}}\,\Lambda\Big)^2
\label{cand_F2}
\;,
\end{equation}
%
whence \eq{relation1} becomes
\begin{eqnarray}
\mathcal{F}_{\alpha\beta}\mathcal{F}_{\mu\nu}
&=&
Q_{\mathcal{C}}^{-1} \Big(\mathcal{C}_{\alpha\beta\mu\nu}
+ \frac{1}{3}Q_{\mathcal{C}}\mathcal{F}^2\mathcal{I}_{\alpha\beta\mu\nu}\Big)
\\
&=&
  \varkappa^{-1} (\mathcal{C}^2)^{-5/6}\Big(\pm  \sqrt{\mathcal{C}^2} -
\sqrt{\frac{32}{3}}\,\Lambda\Big)^2\Big(\mathcal{C}_{\alpha\beta\mu\nu}
\pm
\sqrt{\frac{\mathcal{C}^2}{6}}\,\mathcal{I}_{\alpha\beta\mu\nu}\Big)
\,.
\phantom{xxx}
\label{main_cond}
 \end{eqnarray}
%
Conversely, any  solution of \eq{main_cond} satisfies \eq{relation1}: Indeed,   applying $\mathcal{I}^{\alpha\beta\mu\nu}$ to \eq{main_cond}  yields
\eq{cand_F2},
and thus \eq{relation1} with $Q=Q_{\mathcal{C}}$.
%

Given a KVF $X$ (or its  associated two-form $\mathcal{F}_{\mu\nu}$) such that the associated MST vanishes, the ``right'' sign  in \eq{main_cond} would be determined by the requirement on  $Q_{\mathcal{C}}$  to coincide with $Q_0$.
Here, though, we want to compute $\mathcal{F}_{\mu\nu}$, so a priori it is not clear which sign  in \eq{main_cond} one should take.
Let us therefore investigate to what extent $\mathcal{F}_{\mu\nu}$  is uniquely determined as  solution of \eq{main_cond}, i.e.\
whether it is possible at all that there are different solutions   for both choices    $\pm$.

Let us assume, for contradiction, that \eq{main_cond}  admits a solution for both choices of this sign, $\mathcal{F}_{\mu\nu}$ and $\mathcal{F}'_{\mu\nu}$.
Then, there exist functions $f,f'\ne 0$ such that
\begin{equation}
f \mathcal{F}_{\mu\nu}\mathcal{F}_{\alpha\beta} - f'\mathcal{F}'_{\mu\nu}\mathcal{F}'_{\alpha\beta} \,=\, \mathcal{I}_{\mu\nu\alpha\beta}
\,.
\label{cons0}
\end{equation}
This would imply that any self-dual tensor $\mathcal{W}_{\alpha\beta}$ can be expressed in terms of $\mathcal{F}_{\mu\nu}$ and $\mathcal{F}'_{\mu\nu}$. Indeed,
contraction of \eq{cons0} with $\mathcal{W}^{\mu\nu}$ yields
\begin{equation}
 \mathcal{W}_{\alpha\beta} \,=\, \big(f\mathcal{W}^{\mu\nu}\mathcal{F}_{\mu\nu}\big)\mathcal{F}_{\alpha\beta} - \big( f'\mathcal{W}^{\mu\nu} \mathcal{F}'_{\mu\nu}\big)\mathcal{F}'_{\alpha\beta}
\,.
\end{equation}
Since the space of self-dual two-forms has dimension three, this is a contradiction.
A solution therefore exists at most  for one choice of $\pm$.
Consequently,
given the scale factor $\varkappa\in\mathbb{C}\setminus\{0\}$
(we have already shown
that this is  a gauge factor),
 $\mathcal{F}_{\mu\nu}$ is determined by \eq{main_cond} up to a sign.

A main ingredient for the  algorithm which tests whether a given $\Lambda$-vacuum space-time $(\mcM,g)$ has  a KVF whose associated MST vanishes
 will be to check whether  \eq{main_cond}
%
%
admits a solution $\mathcal{F}_{\alpha\beta}$.
For this purpose let us denote the right-hand side of \eq{main_cond} by $\mathcal{A}_{\alpha\beta\mu\nu}$, so that the equation reads
\begin{equation}
\mathcal{F}_{\alpha\beta}\mathcal{F}_{\mu\nu}
\,=\,
\mathcal{A}_{\alpha\beta\mu\nu}
\;.
\label{F_A_relation}
\end{equation}
This clearly implies ($\mathcal{A}^2:=\mathcal{A}_{\alpha\beta\mu\nu}\mathcal{A}^{\alpha\beta\mu\nu}$)
\begin{eqnarray}
 \mathcal{F}^2\mathcal{F}_{\alpha\beta} \,=\,\mathcal{A}_{\alpha\beta}{}^{\mu\nu}\mathcal{F}_{\mu\nu}
\;, \quad
\mathcal{F}^4\,=\, \mathcal{A}^2
\;, \quad
\mathcal{F}^2
\,=\,
\mathcal{A}_{\alpha\beta}{}^{\alpha\beta}
\,,
 \\
\mathcal{A}_{\alpha\beta\mu\nu}\mathcal{A}^{\mu\nu}{}_{\gamma\delta}
\,=\,
\mathcal{A}_{\mu\nu}{}^{\mu\nu}\mathcal{A}_{\alpha\beta\gamma\delta}
\;.
\label{couple_relations}
\end{eqnarray}
%
%
%
Set $ \mathcal{B}_{\alpha\beta\mu\nu}:= (\mathcal{A}_{\gamma\delta}{}^{\gamma\delta})^{-1} \mathcal{A}_{\alpha\beta\mu\nu}$
(note that $\mathcal{A}_{\gamma\delta}{}^{\gamma\delta}\ne 0 $ in our setting).
It satisfies
\begin{equation}
\mathcal{B}_{\alpha\beta}{}^{\mu\nu}\mathcal{B}_{\mu\nu}{}^{\gamma\delta}
\,=\, \mathcal{B}_{\alpha\beta}{}^{\gamma\delta}
\;, \quad
\mathcal{B}_{\alpha\beta}{}^{\alpha\beta}=1
\;.
\label{in_terms _B}
\end{equation}
We regard $\mathcal{B}_{\alpha\beta}{}^{\mu\nu}$ as an endomorphism  acting on the 3-dimensional space of self-dual two-forms.
 It follows from \eq{in_terms _B} that $\mathcal{B}_{\alpha\beta}{}^{\mu\nu}$ is idempotent with trace 1.
But this implies that  $\mathcal{B}_{\alpha\beta}{}^{\mu\nu}$ can be diagonalized with one eigenvalue equal to $1$ while the two others vanish (in particular it has rank 1).
We denote an eigenvector corresponding to the eigenvalue 1 by $\mathcal{G}_{\mu\nu}$.
Since $\mathcal{G}_{\mu\nu}\ne 0$ there exists  a self-dual two-form  $\mathcal{V}_{\mu\nu}$ such that $\mathcal{V}^{\mu\nu}\mathcal{G}_{\mu\nu}\ne 0$.
Then
\begin{equation}
0 \ne \mathcal{V}^{\mu\nu}\mathcal{G}_{\mu\nu} = \mathcal{V}^{\mu\nu}\mathcal{B}_{\mu\nu}{}^{\alpha\beta}\mathcal{G}_{\alpha\beta}
=  \mathcal{G}^{\mu\nu}\underbrace{\mathcal{B}_{\mu\nu}{}^{\alpha\beta}\mathcal{V}_{\alpha\beta}}_{=\lambda\mathcal{G}_{\mu\nu}}=\lambda \mathcal{G}^2
\;,
\end{equation}
i.e.\ $\mathcal{G}^2\ne 0$ and we normalize it to 1.
Then we have
$\mathcal{B}_{\alpha\beta\mu\nu}= \mathcal{G}_{\alpha\beta} \mathcal{G}_{\mu\nu}$,
and thus \eq{main_cond} with $ \mathcal{F}_{\alpha\beta} =\sqrt{\mathcal{F}^2} \mathcal{G}_{\alpha\beta}$.
(This expresses the fact that any symmetric rank-1  tensor is an outer product of some vector with itself, cf.\ e.g.\ \cite{comon}.)

\begin{lemma}
\label{lemma_C_tensor}
There exists a self-dual tensor $\mathcal{F}_{\alpha\beta}$ with $\mathcal{F}^2\ne 0$ which solves \eq{main_cond} if and only if  
\eq{in_terms _B}
holds, which will be the case
 if and only if the Weyl field\,%
\footnote{
It follows   from the properties of   $\mathcal{C}_{\alpha\beta\mu\nu} $ and $\mathcal{I}_{\alpha\beta\mu\nu}$
that $\mathfrak{C}_{\alpha\beta\gamma\delta}$ is a Weyl~field.
}
\begin{equation}
\mathfrak{C}_{\alpha\beta\gamma\delta}\,:=\,
\mathcal{C}_{\alpha\beta\mu\nu} \mathcal{C}^{\mu\nu}{}_{\gamma\delta}
\mp \sqrt{\frac{\mathcal{C}^2}{6}}\,\mathcal{C}_{\alpha\beta\gamma\delta}
-
\frac{\mathcal{C}^2}{3}\,\mathcal{I}_{\alpha\beta\gamma\delta}\,=\, 0
\;.
\label{item_weyl_0}
\end{equation}
(This will be the case for at most  one choice of $\mp$.)
\end{lemma}


The condition \eq{item_weyl_0} is equivalent to the following two conditions,
\begin{eqnarray}
\mathcal{C}^3:= \mathcal{C}^{\alpha\beta}{}_{\mu\nu} \mathcal{C}^{\mu\nu}{}_{\gamma\delta}\mathcal{C}^{\gamma\delta}{}_{\alpha\beta}
= \pm\frac{(\mathcal{C}^2)^{3/2}}{\sqrt{6}}
\;,
\label{equiv_cond1}
\\
\mathcal{C}_{\alpha\beta\mu\nu} \mathcal{C}^{\mu\nu}{}_{\gamma\delta}
-\frac{\mathcal{C}^3}{\mathcal{C}^2}\mathcal{C}_{\alpha\beta\gamma\delta}
-
\frac{\mathcal{C}^2}{3}\,\mathcal{I}_{\alpha\beta\gamma\delta}= 0
\;.
\label{type_D_cond}
\label{equiv_cond2}
\end{eqnarray}
The latter one characterizes Petrov type D solutions \cite{bel, fms, petrov} and arises from the algebraic classification of the self-dual Weyl tensor
viewed  as a traceless endomorphism on the space of self-dual two-forms.
It  appears as a condition in the Kerr(-NUT)-characterization in \cite{fs} as well.

In particular, \eq{type_D_cond}  recovers the fact that a vacuum space-time with vanishing MST and $\mathcal{F}^2\ne 0$ is of type D.
The type D requirement \eq{type_D_cond} will be  a crucial ingredient of our  characterization. It seems worth emphasizing, though,
  that in itself it  is not strong enough and needs to be complemented by the  additional condition \eq{equiv_cond1}.
As will be shown shortly, \eq{equiv_cond1} makes sure that an eigen-2-form  of the self-dual Weyl tensor supplemented by some trace-term,
 whose existence comes along with the type D property, can be identified with the Killing form of some
(possibly complex)  KVF (cf.\ Theorem~\ref{int_main_result} below).

\subsection{Solving equation \eq{main_cond} explicitly}
\label{sec_solv}
\label{sec_solv_0}

Once it is known, e.g.\ by checking \eq{item_weyl_0}, that \eq{main_cond} admits a solution, there might be the need  to solve this equation explicitly
(in particular if one wishes to derive an explicit expression for the  KVF-candidate).
Here we discuss two  approaches how  a solution of \eq{main_cond} can be constructed algebraically in terms of the self-dual Weyl tensor.

One way of doing that is to solve the equation for $\alpha=\mu$ and $\mu=\nu$ first (without summation),
\begin{equation}
  \mathcal{F}_{\alpha\beta}
 \,=\,
 \frac{1}{2}\Big(\frac{3}{2}\Big)^{1/4}\widetilde\varkappa( \mathcal{C}^2)^{-5/12}\Big(\pm\sqrt{\mathcal{C}^2}-\sqrt{\frac{32}{3}}\,\Lambda\Big)\sqrt{\mathcal{C}_{\alpha\beta\alpha\beta}
\pm \sqrt{\frac{\mathcal{C}^2}{6}}\,\mathcal{I}_{\alpha\beta\alpha\beta}
}
\,,
\end{equation}
where we have set $\widetilde\varkappa:= \pm 2\big(\frac{2}{3}\big)^{1/4} \varkappa^{-1/2}$,
 and to check whether it solves the full set of equations \eq{main_cond} afterwards.

A more elegant way of doing that proceeds as follows:
Denote by $\mathcal{W}_{\alpha\beta}$ any self-dual tensor such that
\begin{equation}
\mathcal{W}^{\mu\nu} \mathcal{C}_{\alpha\beta\mu\nu}
\pm \sqrt{\frac{\mathcal{C}^2}{6}} \mathcal{W}_{\alpha\beta} \,\ne\,  0
\,.
\label{W_ineq}
\end{equation}
The actual sign $\pm$ is determined by \eq{item_weyl_0}.
(If the left-hand-side is zero for all self-dual tensors $\mathcal{W}_{\alpha\beta}$, a solution of  \eq{main_cond} needs to satisfy $\mathcal{F}^2=0$, which
we  exclude in our setting.)
Suppose that \eq{main_cond} admits a solution, then $\mathcal{F}_{\alpha\beta}\mathcal{W}^{\alpha\beta}\ne 0$, i.e.\ there exists
a function $f\ne 0$ such~that
\begin{equation}
\mathcal{F}_{\alpha\beta}
\,=\,
f \,\Big(\mathcal{W}^{\mu\nu}\mathcal{C}_{\alpha\beta\mu\nu}
\pm\sqrt{ \frac{\mathcal{C}^2}{6}}\,\mathcal{W}_{\alpha\beta}\Big)
\,.
\label{cand_F_W}
\end{equation}
%
%
Moreover, $\mathcal{F}_{\alpha\beta}\mathcal{W}^{\alpha\beta}\ne 0$ implies that
\begin{equation}
 \mathcal{W}^{\alpha\beta}\mathcal{W}^{\mu\nu} \mathcal{C}_{\alpha\beta\mu\nu}
\pm  \sqrt{\frac{\mathcal{C}^2}{6}}\, \mathcal{W}^2 \,\ne\,  0
\,.
\label{W_ineq_2}
\end{equation}
We insert \eq{cand_F_W} into \eq{main_cond} and contract with $\mathcal{W}^{\alpha\beta}\mathcal{W}^{\mu\nu}$,
which yields an expression for $f$,
%
\begin{equation}
f
\,=\,
\frac{1}{2} \Big(\frac{3}{2}\Big)^{1/4}\widetilde\varkappa(\mathcal{C}^2)^{-5/12}\Big(\pm \sqrt{\mathcal{C}^2}-\sqrt{\frac{32}{3}}\,\Lambda\Big)\Big(\mathcal{W}^{\alpha\beta}\mathcal{W}^{\mu\nu}\mathcal{C}_{\alpha\beta \mu\nu}
\pm\sqrt{ \frac{\mathcal{C}^2}{6}}\,\mathcal{W}^2\Big)^{-1/2}
\,.
\label{W_expr_f}
\end{equation}
To sum it up, if  \eq{main_cond} admits a solution $\mathcal{F}_{\alpha\beta}$ with $\mathcal{F}^2\ne 0$,
then  there exists a self-dual tensor $\mathcal{W}_{\alpha\beta}$ such that \eq{W_ineq_2} holds, and $\mathcal{F}_{\alpha\beta}$ is given by
\eq{cand_F_W} with $f$ determined  by  \eq{W_expr_f}.
The solution $\mathcal{F}_{\mu\nu}$ is uniquely determined up to a sign.
For given  $\mathcal{C}_{\mu\nu\sigma\rho}$, which can be computed from $g_{\mu\nu}$, and the gauge factor $\varkappa\in\mathbb{C}\setminus\{0\}$,
we can determine  the two-form $\mathcal{F}_{\mu\nu}$ algebraically,
supposing that a solution exists.

Conversely, \eq{cand_F_W}-\eq{W_expr_f}  for some  self-dual tensor $\mathcal{W}_{\alpha\beta}$
provide a candidate $\mathcal{F}_{\alpha\beta}$ to solve \eq{main_cond}; if it does not solve  \eq{main_cond} then there exists  no solution.

\begin{remark}
{\rm
A similar construction for Killing spinors appears in \cite{bk3}.
}
\end{remark}


\subsection{Properties of the candidate field}
\label{sec_KVF_exist}

Guided by the considerations at the beginning of Section~\ref{sec_complex_KVFs}, we define a possibly complex vector field $X$ by \eq{Killing_cand}.
Our aim is  to analyze to what extent   the Killing equation  $\mcL_X g=0$
  is automatically satisfied  in this setting.

First of all let us derive some useful relations which are fulfilled by the
 self-dual two-form  $\mathcal{F}_{\mu\nu}$.
We emphasize  that a priori $\mathcal{F}_{\mu\nu}$ does \emph{not} satisfy most of
the above relations,
 namely \eq{rln_X_F}, \eq{rln_nablaF}-\eq{rln_chi} and
\eq{rln_nablaF_2}-\eq{rln_nablaF2_2},
whereas \eq{self_dual} and \eq{self_one}-\eq{useful_first}, which follow from the self-duality of $\mathcal{F}_{\mu\nu}$, \emph{do} hold.

%
We compute the divergence of \eq{relation1},
\begin{eqnarray}
\mathcal{F}^{-2}\frac{Q\mathcal{F}^2   +20\Lambda }{Q\mathcal{F}^2 +2\Lambda} \mathcal{F}_{\alpha\beta}\mathcal{F}_{\mu}{}^{\nu} \nabla_{\nu}\mathcal{F}^2
-4\mathcal{F}_{\mu}{}^{\nu} \nabla_{\nu} \mathcal{F}_{\alpha\beta}
-4 \mathcal{F}_{\alpha\beta}\nabla_{\nu} \mathcal{F}_{\mu}{}^{\nu}
\nonumber
&&
\\
+ \frac{Q\mathcal{F}^2   -4 \Lambda }{Q\mathcal{F}^2 +2\Lambda} \nabla_{\nu}\mathcal{F}^2
\mathcal{I}_{\alpha\beta\mu}{}^{\nu}
&=&  0
\,, \phantom{xxx}
\label{relation_divergence}
 \end{eqnarray}
where we have used \eq{relation2}.
Contraction with $\mathcal{F}^{\alpha\beta}$ yields with  \eq{Killing_cand},
\begin{equation}
 \nabla_{\nu} \mathcal{F}_{\mu}{}^{\nu} \,=\,  \Lambda X_{\mu}
\,.
\label{relation_divF}
\end{equation}
Contracting  \eq{relation_divergence} with  $\mathcal{F}^{\mu}{}_{\kappa}$ and using  \eq{relation_F2} and \eq{relation_divF}, one is led   to
%
\begin{equation}
\mathcal{F}^2 \nabla_{\kappa} \mathcal{F}_{\alpha\beta}
- \frac{1}{4}\frac{Q\mathcal{F}^2   +8\Lambda }{Q\mathcal{F}^2 +2\Lambda} \mathcal{F}_{\alpha\beta} \nabla_{\kappa}\mathcal{F}^2
+ \frac{Q\mathcal{F}^2   -4 \Lambda }{Q\mathcal{F}^2 +2\Lambda}\mathcal{F}_{\kappa\mu} \nabla_{\nu}\mathcal{F}^2
\mathcal{I}_{\alpha\beta}{}^{\mu\nu}
\,=\, 0
\,.
\label{relation_nabla}
 \end{equation}
Let us  determine the symmetric part of  the divergence of \eq{relation_nabla}. A somewhat lenghty calculation which uses
\eq{relation_divF}, \eq{relation2}, \eq{Killing_cand}, \eq{relation_nabla}, the self-duality of $\mathcal{F}_{\mu\nu}$,
and  the $\Lambda$-vacuum Einstein equations    reveals that
\begin{eqnarray}
0
&=&
 \nabla^{\beta}\mathcal{F}^2 \nabla_{(\kappa} \mathcal{F}_{\alpha)\beta}
 + \Lambda \mathcal{F}^2  \nabla_{(\kappa}X_{\alpha)}
 +\mathcal{F}^2 R_{\gamma(\kappa}\mathcal{F}_{\alpha)}{}^{\gamma}
- \frac{1}{2} \mathcal{F}_{(\alpha}{}^{\beta} \nabla_{\kappa)} \nabla_{\beta}\mathcal{F}^2
\nonumber
\\
&&
+ \frac{3}{2}(\mathcal{F}^2 \nabla_{\beta}Q +Q\nabla_{\beta}\mathcal{F}^2)  \frac{\Lambda }{(Q\mathcal{F}^2 +2\Lambda)^2} \mathcal{F}_{(\alpha}{}^{\beta} \nabla_{\kappa)}\mathcal{F}^2
\nonumber
\\
&&
- \frac{1}{4}\Lambda \frac{Q\mathcal{F}^2   +8\Lambda }{Q\mathcal{F}^2 +2\Lambda}X_{(\alpha} \nabla_{\kappa)}\mathcal{F}^2
+ \frac{Q\mathcal{F}^2   -4 \Lambda }{Q\mathcal{F}^2 +2\Lambda} \nabla_{\beta}\mathcal{F}_{(\kappa|\mu} \nabla_{\nu|}\mathcal{F}^2
\mathcal{I}_{\alpha)}{}^{\beta\mu\nu}
\nonumber
\\
&&
+6 ( \mathcal{F}^2  \nabla_{\beta}Q+Q  \nabla_{\beta}\mathcal{F}^2) \frac{\Lambda }{(Q\mathcal{F}^2 +2\Lambda)^2}\mathcal{F}_{(\kappa|\mu} \nabla_{\nu|}\mathcal{F}^2
\mathcal{I}_{\alpha)}{}^{\beta\mu\nu}
\\
&=&
 - \Lambda X_{(\alpha} \nabla_{\kappa)}\mathcal{F}^{2}\Big( 1
+ \frac{1}{4} \frac{Q\mathcal{F}^2   +8\Lambda }{Q\mathcal{F}^2 +2\Lambda}
 +\frac{3}{4} \frac{Q\mathcal{F}^2   -4 \Lambda }{(Q\mathcal{F}^2 +2\Lambda)^2} Q \mathcal{F}^2
 \Big)
\nonumber
\\
&& +   \frac{Q \mathcal{F}^2  +5\Lambda  }{Q \mathcal{F}^2  +2\Lambda } \nabla_{(\kappa}  \mathcal{F}_{\alpha)\beta} \nabla^{\beta}\mathcal{F}^2
 -\frac{1}{2} \frac{Q \mathcal{F}^2   -4  \Lambda}{Q \mathcal{F}^2  +2\Lambda } \mathcal{F}_{(\alpha}{}^{\beta} \nabla_{\kappa)} \nabla_{\beta}\mathcal{F}^2
\nonumber
\\
&&
+ \frac{Q\mathcal{F}^2   -4 \Lambda }{Q\mathcal{F}^2 +2\Lambda} \nabla_{\beta}\mathcal{F}_{(\kappa|\mu} \nabla_{\nu|}\mathcal{F}^2
\mathcal{I}_{\alpha)}{}^{\beta\mu\nu}
\\
&=&
\frac{1}{24}(5(Q\mathcal{F}^2)^2 - 4\Lambda Q\mathcal{F}^2  + 8\Lambda^2)\frac{Q \mathcal{F}^2  -4\Lambda }{(Q\mathcal{F}^2 +2\Lambda)^2} X_{(\alpha} \nabla_{\kappa)}\mathcal{F}^2
\nonumber
\\
&&
 -\frac{1}{2} \frac{Q \mathcal{F}^2   -4  \Lambda}{Q \mathcal{F}^2  +2\Lambda } \mathcal{F}_{(\alpha}{}^{\beta} \nabla_{\kappa)} \nabla_{\beta}\mathcal{F}^2
\,,
 \end{eqnarray}
whence
%
\begin{equation}
\mathcal{F}_{(\alpha}{}^{\beta} \nabla_{\kappa)} \nabla_{\beta}\mathcal{F}^2
\,=\,
\frac{1}{12}\frac{5(Q\mathcal{F}^2)^2 - 4\Lambda Q\mathcal{F}^2  + 8\Lambda^2 }{Q\mathcal{F}^2 +2\Lambda} X_{(\alpha} \nabla_{\kappa)}\mathcal{F}^2
 \,.
\label{rln_second_order}
 \end{equation}

Eventually, we consider the Killing equation for a vector field $X$ which is given by  \eq{Killing_cand}. Using  \eq{Killing_cand}, \eq{relation2} and  \eq{relation_nabla},
we find for its covariant derivative,
\begin{eqnarray}
\nabla_{\mu} X_{\nu}
&=&-\mathcal{F}^2 \nabla_{\mu} Q\frac{1}{Q \mathcal{F}^2  +2\Lambda}  X_{\nu}
- \frac{2Q\mathcal{F}^2+2\Lambda}{Q \mathcal{F}^2  +2\Lambda}\mathcal{F}^{-2}  X_{\nu}\nabla_{\mu}\mathcal{F}^2
\nonumber
\\
&& + \frac{3}{Q \mathcal{F}^2  +2\Lambda}  \mathcal{F}^{-2}  (\nabla_{\mu}\mathcal{F}_{\nu\alpha} \nabla^{\alpha}\mathcal{F}^2
 + \mathcal{F}_{\nu}{}^{\alpha} \nabla_{\mu}\nabla_{\alpha}\mathcal{F}^2)
\\
&=&\Big(  -\frac{5}{4}
+ 3\Lambda\frac{2Q\mathcal{F}^2+ \Lambda }{(Q\mathcal{F}^2 +2\Lambda)^2}  \Big) \mathcal{F}^{-2}X_{\nu}\nabla_{\mu}\mathcal{F}^2
\nonumber
\\
&&
+  \frac{1}{2} \frac{Q\mathcal{F}^2   -4 \Lambda }{Q\mathcal{F}^2 +2\Lambda} \mathcal{F}^{-2}X_{[\mu} \nabla_{\nu]}\mathcal{F}^2
 + \frac{3}{Q \mathcal{F}^2  +2\Lambda} \mathcal{F}^{-2} \mathcal{F}_{\nu}{}^{\alpha} \nabla_{\mu}\nabla_{\alpha}\mathcal{F}^2
\nonumber
\\
&&
-  \frac{3}{4} \frac{Q\mathcal{F}^2   -4 \Lambda }{(Q\mathcal{F}^2 +2\Lambda)^2} \mathcal{F}^{-4} \mathcal{F}_{\mu\nu} \nabla_{\alpha}\mathcal{F}^2 \nabla^{\alpha}\mathcal{F}^2
\,.
\label{nabla_Killing_cand}
\end{eqnarray}
With \eq{rln_second_order} its symmetric part reads
\begin{eqnarray}
\mathcal{F}^2\nabla_{(\mu} X_{\nu)}
&\hspace{-0.5em}=&\hspace{-0.5em}
 \frac{3}{ Q \mathcal{F}^2  +2\Lambda } \mathcal{F}_{(\mu}{}^{\alpha} \nabla_{\nu)}\nabla_{\alpha}\mathcal{F}^2
-\Big( \frac{5}{4}
-\frac{ 3\Lambda(2Q\mathcal{F}^2 +\Lambda)  }{(Q\mathcal{F}^2 +2\Lambda)^2} \Big) X_{(\mu} \nabla_{\nu)}\mathcal{F}^2
\nonumber
\\
&\hspace{-0.5em}=&\hspace{-0.5em}
0
\,,
\end{eqnarray}
i.e.\ $X$ is a KVF with norm
 \begin{equation}
 |X|^2  \,=\, \frac{9}{4}\mathcal{F}^{-2}(Q \mathcal{F}^2  +2\Lambda)^{-2}  \nabla_{\alpha}\mathcal{F}^2\nabla^{\alpha}\mathcal{F}^2
\,.
\end{equation}

We have assumed the existence of a two-form  $\mathcal{F}_{\mu\nu}$ such that the associated ``MST''%
\footnote{
The quotation marks are to emphasize that this is not a proper MST as long as we do not know whether $\mathcal{F}_{\mu\nu}$
arises from a KVF.
}
 vanishes and we have derived a vector field $X^{\mu}$ from  $\mathcal{F}_{\mu\nu}$
which satisfies the Killing equation.
However, we do not now yet whether $\mathcal{F}_{\mu\nu}$ and $X^{\mu}$ are related in the desired way, by which we mean that
\begin{equation}
\mathcal{F}_{\alpha\beta} \,=\,
 \nabla_{\alpha}X_{\beta} + i (\nabla_{\alpha}X_{\beta})^{\star} \,\equiv\,
2 \mathcal{I}_{\alpha\beta}{}^{\mu\nu}\nabla_{\mu}X_{\nu}
\,.
\label{right_F}
\end{equation}
We need to make  sure that this relation is fulfilled to conclude that the MST associated to the KVF $X$ vanishes.

For this, let us consider the skew-symmetric part of \eq{nabla_Killing_cand},
\begin{eqnarray}
\nabla_{[\mu} X_{\nu]}
&=&\frac{7(Q\mathcal{F}^2)^2-8Q\mathcal{F}^2\Lambda -8 \Lambda^2 }{4(Q\mathcal{F}^2 +2\Lambda)^2} \mathcal{F}^{-2} X_{[\mu}\nabla_{\nu]}\mathcal{F}^2
 - \frac{3 \mathcal{F}^{-2}}{Q \mathcal{F}^2  +2\Lambda} \mathcal{F}_{[\mu}{}^{\alpha} \nabla_{\nu]}\nabla_{\alpha}\mathcal{F}^2
\nonumber
\\
&&
-  \frac{3}{4} \frac{Q\mathcal{F}^2   -4 \Lambda }{(Q\mathcal{F}^2 +2\Lambda)^2}\mathcal{F}^{-4}  \mathcal{F}_{\mu\nu} \nabla_{\alpha}\mathcal{F}^2 \nabla^{\alpha}\mathcal{F}^2
\,.
\end{eqnarray}
We apply $\mathcal{I}_{\alpha\beta}{}^{\mu\nu}$,
\begin{eqnarray}
\mathcal{I}_{\alpha\beta}{}^{\mu\nu}\nabla_{\mu} X_{\nu}
\hspace{-0.7em}
&=&
\hspace{-0.7em}
\frac{3}{4}\frac{7(Q\mathcal{F}^2)^2-8Q\mathcal{F}^2\Lambda -8 \Lambda^2 }{(Q\mathcal{F}^2 +2\Lambda)^3}\mathcal{F}^{-4}\mathcal{I}_{\alpha\beta}{}^{\mu(\nu}  \mathcal{F}_{\mu}{}^{\gamma)} \nabla_{\nu}\mathcal{F}^2\nabla_{\gamma}\mathcal{F}^2
\nonumber
\\
&&
\hspace{-0.7em}
 - \frac{3}{Q \mathcal{F}^2  +2\Lambda} \mathcal{F}^{-2}\mathcal{I}_{\alpha\beta}{}^{\mu(\nu} \mathcal{F}_{\mu}{}^{\gamma)} \nabla_{\nu}\nabla_{\gamma}\mathcal{F}^2
\nonumber
\\
&&
\hspace{-0.7em}
-  \frac{3}{4} \frac{Q\mathcal{F}^2   -4 \Lambda }{(Q\mathcal{F}^2 +2\Lambda)^2}  \mathcal{F}^{-4}\mathcal{F}_{\alpha\beta} \nabla^{\gamma}\mathcal{F}^2 \nabla_{\gamma}\mathcal{F}^2
\\
&=&
\hspace{-0.7em}
 - \frac{3}{4}\frac{\mathcal{F}^{-2}}{Q \mathcal{F}^2  +2\Lambda}\Big(\Box\mathcal{F}^2
-\frac{3}{4}\frac{(Q\mathcal{F}^2)^2  +8 \Lambda^2
 }{(Q\mathcal{F}^2 +2\Lambda)^2}\mathcal{F}^{-2} \nabla^{\gamma}\mathcal{F}^2\nabla_{\gamma}\mathcal{F}^2
\Big) \mathcal{F}_{\alpha\beta}
\,.
\phantom{xxxx}
\label{towards_F_X}
\end{eqnarray}
%

We need to derive an expression for $\Box\mathcal{F}^2$.
Contraction of \eq{towards_F_X} with $\mathcal{F}^{\alpha\beta}$ gives the useful relation
\begin{equation}
\mathcal{F}^{\mu\nu}\nabla_{\mu} X_{\nu}
=
-\frac{3}{4}\frac{1}{Q \mathcal{F}^2  +2\Lambda} \Big( \Box \mathcal{F}^2
-\frac{3}{4}\frac{(Q\mathcal{F}^2)^2+8 \Lambda^2 }{(Q\mathcal{F}^2 +2\Lambda)^2}
\mathcal{F}^{-2} \nabla^{\gamma}\mathcal{F}^2 \nabla_{\gamma}\mathcal{F}^2
\Big)
\,.
\label{another_useful_rln}
\end{equation}
Next, let us  determine the skew-symmetric part of  the divergence of \eq{relation_nabla}.
Another  lengthy calculation which makes use of
\eq{relation_divF}, \eq{relation1}, \eq{relation2},
the self-duality of $\mathcal{F}_{\mu\nu}$, and the vacuum Einstein equations
reveals that
%
\begin{eqnarray}
 0
 &=&
\nabla^{\beta}\mathcal{F}^2 \nabla_{[\kappa} \mathcal{F}_{\alpha]\beta}
+ \mathcal{F}^2 \nabla_{[\kappa}\nabla_{|\beta|} \mathcal{F}_{\alpha]}{}^{\beta}
 -\frac{1}{2}\mathcal{F}^2  R_{\alpha\kappa\beta\gamma}\mathcal{F}^{\beta\gamma} + \mathcal{F}^2  R_{\gamma[\kappa}\mathcal{F}_{\alpha]}{}^{\gamma}
\nonumber
\\
&&
+ \frac{3}{2}(2Q\nabla_{\beta}\mathcal{F}^2 + \mathcal{F}^2\nabla_{\beta}Q )\frac{\Lambda }{(Q\mathcal{F}^2 +2\Lambda)^2} \mathcal{F}_{[\alpha}{}^{\beta} \nabla_{\kappa]}\mathcal{F}^2
\nonumber
\\
&&
+ \frac{1}{4}\frac{Q\mathcal{F}^2   +8\Lambda }{Q\mathcal{F}^2 +2\Lambda} \nabla^{\beta}\mathcal{F}_{\beta[\alpha} \nabla_{\kappa]}\mathcal{F}^2
- \frac{3}{2} Q \frac{ \Lambda }{(Q\mathcal{F}^2 +2\Lambda)^2} \mathcal{F}_{\alpha \kappa}\nabla_{\beta}\mathcal{F}^2 \nabla^{\beta}\mathcal{F}^2
\nonumber
\\
&&
- 3\frac{ \Lambda }{Q\mathcal{F}^2 +2\Lambda} \mathcal{F}_{[\alpha}{}^{\beta} \nabla_{\kappa]}\nabla_{\beta}\mathcal{F}^2
-6   \mathcal{F}^2\nabla_{\beta}Q \frac{ \Lambda }{(Q\mathcal{F}^2 +2\Lambda)^2} \nabla_{\nu}\mathcal{F}^2 \mathcal{F}_{\mu[\kappa}\mathcal{I}_{\alpha]}{}^{\beta\mu\nu}
\nonumber
\\
&&
- \frac{Q\mathcal{F}^2   -4 \Lambda }{Q\mathcal{F}^2 +2\Lambda}\nabla_{\nu}\mathcal{F}^2\nabla_{\beta}\mathcal{F}_{\mu[\kappa}  \mathcal{I}_{\alpha]}{}^{\beta\mu\nu}
- \frac{1}{4} \frac{Q\mathcal{F}^2   -4 \Lambda }{Q\mathcal{F}^2 +2\Lambda}\mathcal{F}_{\alpha\kappa} \Box \mathcal{F}^2
\\
 &=&
\nabla^{\beta}\mathcal{F}^2 \nabla_{[\kappa} \mathcal{F}_{\alpha]\beta}
- \Lambda \mathcal{F}^2 \nabla_{[\alpha}X_{\kappa]}
 -\frac{1}{6}(Q\mathcal{F}^2-4\Lambda)\mathcal{F}^2 \mathcal{F}_{\alpha\kappa}
\nonumber
\\
&&
+ \frac{9}{4}\Lambda
 \frac{ Q\mathcal{F}^2   -4\Lambda}{(Q\mathcal{F}^2 +2\Lambda)^3}Q\mathcal{F}_{[\alpha}{}^{\beta} \nabla_{\kappa]}\mathcal{F}^2\nabla_{\beta}\mathcal{F}^2
- 3\frac{ \Lambda }{Q\mathcal{F}^2 +2\Lambda} \mathcal{F}_{[\alpha}{}^{\beta} \nabla_{\kappa]}\nabla_{\beta}\mathcal{F}^2
\nonumber
\\
&&
- \frac{\Lambda}{4}\frac{Q\mathcal{F}^2   +8\Lambda }{Q\mathcal{F}^2 +2\Lambda}X_{[\alpha} \nabla_{\kappa]}\mathcal{F}^2
- \frac{9}{8}\Lambda  \frac{ Q\mathcal{F}^2   -4\Lambda  }{(Q\mathcal{F}^2 +2\Lambda)^3} Q\nabla^{\beta}\mathcal{F}^2\nabla_{\beta}\mathcal{F}^2 \mathcal{F}_{\alpha\kappa}
\nonumber
\\
&&
- \frac{Q\mathcal{F}^2   -4 \Lambda }{Q\mathcal{F}^2 +2\Lambda}\nabla_{\nu}\mathcal{F}^2\nabla_{\beta}\mathcal{F}_{\mu[\kappa}  \mathcal{I}_{\alpha]}{}^{\beta\mu\nu}
- \frac{1}{4} \frac{Q\mathcal{F}^2   -4 \Lambda }{Q\mathcal{F}^2 +2\Lambda}\mathcal{F}_{\alpha\kappa} \Box \mathcal{F}^2
\,.
\end{eqnarray}
Contraction with $\mathcal{F}^{\alpha\kappa}$ yields with \eq{another_useful_rln}, \eq{relation_F2},
\eq{Killing_cand}, \eq{relation_nabla}
and \eq{important_relation},
\begin{eqnarray}
0 &=&
- \frac{1}{4} \frac{Q\mathcal{F}^2   -4 \Lambda }{Q\mathcal{F}^2 +2\Lambda}\mathcal{F}^2 \Box \mathcal{F}^2
+\mathcal{F}^{\alpha\kappa}\nabla^{\beta}\mathcal{F}^2 \nabla_{\kappa} \mathcal{F}_{\alpha\beta}
 -\frac{1}{6}(Q\mathcal{F}^2-4\Lambda)\mathcal{F}^4
\nonumber
\\
&&
- \frac{3}{16}\Lambda
 \frac{ 7(Q\mathcal{F}^2)^2   -2Q\mathcal{F}^2  \Lambda + 40\Lambda^2
}{(Q\mathcal{F}^2 +2\Lambda)^3}\nabla^{\beta}\mathcal{F}^2\nabla_{\beta}\mathcal{F}^2
\nonumber
\\
&&
- \frac{Q\mathcal{F}^2   -4 \Lambda }{Q\mathcal{F}^2 +2\Lambda}\mathcal{F}^{\alpha\kappa}\nabla_{\nu}\mathcal{F}^2\nabla_{\beta}\mathcal{F}_{\mu[\kappa}  \mathcal{I}_{\alpha]}{}^{\beta\mu\nu}
\\
&=& - \frac{1}{4}\mathcal{F}^2 \frac{Q\mathcal{F}^2-4\Lambda}{Q\mathcal{F}^2 +2\Lambda}  \Big(
\Box \mathcal{F}^2
 +\frac{2}{3}\mathcal{F}^2(Q\mathcal{F}^2 +2\Lambda)
\nonumber
\\
&&
 \phantom{xxxxxxxxxxxxxxxx} -\frac{3}{4}\frac{(Q\mathcal{F}^2)^2 +8\Lambda^2
}{(Q\mathcal{F}^2 +2\Lambda)^2}\mathcal{F}^{-2} \nabla^{\beta}\mathcal{F}^2\nabla_{\beta}\mathcal{F}^2
\Big)
\,.
\end{eqnarray}
Because of  \eq{relation3} it follows that
\begin{equation}
\Box \mathcal{F}^2
 +\frac{2}{3}\mathcal{F}^2(Q\mathcal{F}^2 +2\Lambda)
 -\frac{3}{4}\frac{(Q\mathcal{F}^2)^2 +8\Lambda^2
}{(Q\mathcal{F}^2 +2\Lambda)^2} \mathcal{F}^{-2}\nabla^{\beta}\mathcal{F}^2\nabla_{\beta}\mathcal{F}^2
\,=\, 0
\,.
\label{eqn_wave_F}
\end{equation}
In fact, this is the only step where the assumption $Q\mathcal{F}^2-4\Lambda\ne 0 $of  \eq{relation3} enters.

We insert \eq{eqn_wave_F}  into \eq{towards_F_X} to conclude that \eq{right_F} holds,
%
%
i.e.\ $\mathcal{F}_{\mu\nu}$ and $X^{\mu}$ are automatically related in the desired way.
For  $X^{\mu}=0$ \eq{right_F} would imply  $\mathcal{F}^2=0$, which is excluded in our setting.

We have established the following result:
\begin{theorem}
\label{thm_X_in_F}
Assume that a $\Lambda$-vacuum space-time $(\mcM,g)$ admits a self-dual two-form $\mathcal{F}_{\mu\nu}$
and a function $Q:\mcM\rightarrow \mathbb{C}$  such that \eq{relation1}-\eq{relation3} hold. 
Define a (possibly complex) vector field $X$ by
 \begin{equation}
  \label{10VI15.1}
 X^{\mu} \,=\, 3(Q \mathcal{F}^2  +2\Lambda)^{-1}   \mathcal{F}^{\mu\alpha} \nabla_{\alpha}\log\mathcal{F}^2
\;.
\end{equation}
Then:
\begin{enumerate}
\item[(i)]   $X$ is a non-zero vector field which satisfies the Killing equation $\mcL_X g=0$ (in particular its real and imaginary part are
real KVFs, though one of them might be trivial).
\item[(ii)]  The MST associated to $X$ vanishes.
\end{enumerate}
\end{theorem}

\begin{remark}
{\rm
In Theorem~\ref{thm_X_in_F} the KVF might be complex. In \cite{mpss} a complete classification of $\Lambda$-vacuum space-times is given
which admit a \emph{real} KVF such that the associated MST vanishes.
It would be interesting to study the class of space-times which admit a complex KVF such that the associated MST vanishes.
}
\end{remark}


Using \eq{cand_Q}-\eq{cand_F2},
the expression \eq{10VI15.1} for $X$ becomes\,%
\footnote{
For the expression on the right-hand side of \eq{10VI15.1} to be well-defined the condition $Q\mathcal{F}^2+2 \Lambda\ne 0$
(i.e.\ $\pm\sqrt{\mathcal{C}^2} + \sqrt{\frac{8}{3}}\,\Lambda \ne 0$) needs to be satisfied.
However, the final expression \eq{Killing_cand2} is well-defined even if equality holds.
So an interesting question arises whether a vanishing MST implies that $X$ needs to be of the form  \eq{Killing_cand2} also for
 $Q\mathcal{F}^2+2 \Lambda= 0$.
}
%
\begin{eqnarray}
 X^{\mu}
&=&
 \sqrt{6}\Big(\pm
\sqrt{\mathcal{C}^2} +\sqrt{\frac{8}{3}}\Lambda\Big)^{-1}  \mathcal{F}^{\mu\alpha} \nabla_{\alpha}\log
\frac{\Big( \pm\sqrt{\mathcal{C}^2}
- \sqrt{\frac{32}{3}}\,\Lambda\Big)^2}{(\mathcal{C}^2)^{1/3}}
\phantom{xx}
\\
&=&
\sqrt{\frac{8}{3}}\Big( \pm
\sqrt{\mathcal{C}^2} -\sqrt{\frac{32}{3}}\,\Lambda\Big)^{-1}
\mathcal{F}^{\mu\alpha} \nabla_{\alpha} \log\mathcal{C}^2
\;,
\label{Killing_cand2}
\\
\Longrightarrow \quad   |X|^2 &=&\pm  \frac{1}{4}\widetilde\varkappa^2
 (\mathcal{C}^2)^{-7/3}\nabla^{\alpha} \mathcal{C}^2\nabla_{\alpha} \mathcal{C}^2
\,.
\label{norm_candidate_field}
\end{eqnarray}
%
%
It follows from Lemma~\ref{relation_Q_diffQ}, Lemma~\ref{lemma_C_tensor} and Theorem~\ref{thm_X_in_F} that $X$ is a non-trivial KVF w.r.t.\ which the MST
vanishes whenever \eq{item_weyl_0} holds.
Then the procedure described in Section~\ref{sec_solv_0} must yield a solution $\mathcal{F}_{\mu\nu}$.
More specifically, if $\mathcal{W}_{\alpha\beta}$ is a self-dual tensor which satisfies \eq{W_ineq_2} then \eq{cand_F_W} together with \eq{W_expr_f}
solve \eq{main_cond} so that
\eq{Killing_cand2} provides a KVF whose associated MST vanishes.
It reads
As above, we set  $\widetilde\varkappa =  \pm2\big(\frac{2}{3}\big)^{1/4} \varkappa^{-1/2}$, then
\begin{equation}
 X^{\alpha} \,=\,
\frac{\big(\frac{2}{3}\big)^{1/4}\widetilde\varkappa
\Big(\mathcal{C}^{\alpha\beta}{}_{\mu\nu} \mathcal{W}^{\mu\nu}
\pm\sqrt{ \frac{\mathcal{C}^2}{6}}\,\mathcal{W}^{\alpha\beta}\Big) \nabla_{\beta}\log \mathcal{C}^2
}
{(\mathcal{C}^2)^{5/12} \sqrt{\mathcal{W}^{\gamma\delta}\mathcal{W}^{\sigma\rho}\mathcal{C}_{\gamma\delta\sigma\rho}
\pm\sqrt{ \frac{\mathcal{C}^2}{6}}\,\mathcal{W}^2}}
\;.
\label{final_eqn_KVF0}
\end{equation}
It is convenient to employ \eq{item_weyl_0} to simplify the computation of the KVF.
We have
\begin{equation}
\Big|  \mathcal{C}^{\alpha\beta}{}_{\mu\nu} \mathcal{W}^{\mu\nu}
\pm\sqrt{ \frac{\mathcal{C}^2}{6}}\,\mathcal{W}^{\alpha\beta}\Big|^2
\,=\, \sqrt{\frac{3}{2}\mathcal{C}^2} \Big(\mathcal{W}^{\gamma\delta}\mathcal{W}^{\sigma\rho}\mathcal{C}_{\gamma\delta\sigma\rho}
\pm\sqrt{ \frac{\mathcal{C}^2}{6}}\,\mathcal{W}^2
\Big)
\;.
\end{equation}
Let us assume that $\mathcal{W}_{\alpha\beta}$ has been normalized in such a way that this norm is $1$.
Then \eq{final_eqn_KVF0} becomes \eq{final_eqn_KVF} below.

\subsection{Uniqueness of KVFs with vanishing MST}
\label{sec_uniqueness}

We have shown so far that a solution $(\mathcal{F}_{\mu\nu}, Q=Q_{\mathcal{C}})$
  of \eq{relation1}
yields a (possibly complex) non-trivial KVF $X$, given by \eq{final_eqn_KVF0},  such that the associated MST vanishes.
Conversely, as the following Proposition shows,   $X$ is, up to rescaling, the only KVF with this property.
%
%
This reflects the fact
that the MST vanishes w.r.t.\  the KVF $X$ if and only if it vanishes w.r.t.\ $\lambda X$, $\lambda\in\mathbb{C}\setminus \{0\}$, and that
there is at most one such 1-parameter family with this property:


%
\begin{proposition}
\label{prop_uniqueness}
 Let $(\mcM,g)$ be a  smooth $(3+1)$-dimensional $\Lambda$-vacuum space-time which satisfies \eq{ineqs_C2_new}
and which admits a (possibly complex)
KVF $X$  such the associated MST vanishes for some function $Q$.
Then $X$ is, up to scaling with some non-zero complex constant,  the only  KVF with this property. 
\end{proposition}
\begin{remark}
{\rm
If one finds a properly complex KVF which is not a real one multiplied by some complex constant, such that the associated MST vanishes, it follows from Proposition~\ref{prop_uniqueness} that there exists no real KVF with this property.
}
\end{remark}
\begin{proof}
Let us assume that there are two KVFs $X$ and $X'$ such that the associated MST vanishes.
Rescaling one of them, if necessary, we  have by  Proposition~\ref{prop_rln_Q2} $Q=Q_{\mathcal{C}}=Q'$.
The associated $\mathcal{F}_{\alpha\beta}$ and $\mathcal{F}'_{\alpha\beta}$ both  satisfy \eq{main_cond}.
The  argument in Section~\ref{sec_petrov} which
 establishes the uniqueness of solutions thereof shows that the right-hand sides are identical, whence $\mathcal{F}_{\alpha\beta}=\pm \mathcal{F}'_{\alpha\beta}$.
Equation \eq{Killing_cand2} then shows that $X=\pm X'$.
\qed
\end{proof}

\subsection{Algorithmic  characterization of space-times which admit a complex KVF w.r.t.\ which the MST vanishes}
\label{section_alg_charact}

We end up with our first main result:
\begin{theorem}
\label{int_main_result}
Let $(\mcM,g)$ be a  smooth $(3+1)$-dimensional $\Lambda$-vacuum space-time
containing an open dense set on which\,%
\footnote{
\label{footnote_weaker}
As in Section~\ref{sec_yet_another} the latter two may be replaced by
$\pm \sqrt{\mathcal{C}^2} - \sqrt{\frac{32}{3}}\,\Lambda\ne 0$ and
$ \pm\sqrt{\mathcal{C}^2} + \sqrt{\frac{8}{3}} \Lambda \ne 0$, respectively;
moreover, they
are allowed to be violated on surfaces of co-dimension $\geq 1$, cf.\ footnote~\ref{footnote_codimension}.
}
\begin{equation}
\mathcal{C}^2 \,\ne\, 0\;, \quad  \mathcal{C}^2 - \frac{32}{3}\Lambda^2\,\ne\, 0\;,
\quad   \mathcal{C}^2 -\frac{8}{3} \Lambda^2 \,\ne\, 0
\;.
\label{all_ineq_C2}
\end{equation}
Then, if and only if
\begin{itemize}
\item  the self-dual Weyl tensor corrected by a trace-term,
$\mathcal{W}_{\alpha\beta\mu\nu} := \mathcal{C}_{\alpha\beta\mu\nu}
\pm \sqrt{\frac{\mathcal{C}^2}{6}}\,\mathcal{I}_{\alpha\beta\mu\nu}$, satisfies $\mathcal{W}=\omega\otimes\omega$ for some two-form $\omega$,
\end{itemize}
or, equivalently,
\begin{itemize}
\item
the Weyl field
$\mathfrak{C}_{\alpha\beta\gamma\delta}\equiv  \mathcal{C}_{\alpha\beta\mu\nu} \mathcal{C}^{\mu\nu}{}_{\gamma\delta}
\mp \sqrt{\frac{\mathcal{C}^2}{6}}\,\mathcal{C}_{\alpha\beta\gamma\delta}
-
\frac{\mathcal{C}^2}{3}\,\mathcal{I}_{\alpha\beta\gamma\delta} $
vanishes
(the conditions will be satisfied for at most one sign $\mp$),
\end{itemize}
there exists a non-trivial, possibly complex KVF $X$ such that the associated MST vanishes for some function $Q$.
It is, up to rescaling, the only KVF with this property, and can be computed as follows:
Let $\mathcal{W}_{\mu\nu}$ be a self-dual tensor  with
$|  \mathcal{C}^{\alpha\beta}{}_{\mu\nu} \mathcal{W}^{\mu\nu}
\pm\sqrt{ \frac{\mathcal{C}^2}{6}}\,\mathcal{W}^{\alpha\beta}|^2=1$,
then
\begin{equation}
 X^{\alpha} \,=\,
\widetilde\varkappa (\mathcal{C}^2)^{-1/6}
\Big(\mathcal{C}^{\alpha\beta}{}_{\mu\nu} \mathcal{W}^{\mu\nu}
\pm\sqrt{ \frac{\mathcal{C}^2}{6}}\,\mathcal{W}^{\alpha\beta}\Big) \nabla_{\beta}\log \mathcal{C}^2
\;.
\label{final_eqn_KVF}
\end{equation}
\end{theorem}

\begin{remark}
{\rm
Compared to the MST the definition of the tensor $\mathfrak{C}_{\alpha\beta\gamma\delta}$ does not rely on the existence of a KVF.
It thus  may provide a tool to construct a geometric invariant which measures  the deviation of a given $\Lambda$-vacuum space-time,
without any symmetries, to one  which admits
a (complex or real)  KVF with vanishing MST, and ultimately also to a member of the Kerr-(A)dS family.%
\footnote{Scalar quantities, so-called  ``quality factors'', which measure the deviation from Kerr in the setting of \emph{stationary} space-times
have been constructed in \cite{senovilla}.
}
Some care is needed, though, due to the sign ambiguity in its definition: If the tensor $\mathfrak{C}_{\alpha\beta\gamma\delta}$ vanishes for $\mathcal{C}^2\ne 0$, it vanishes either for $+$ or for $-$, so  there is a preferred choice.
If it does not vanish such a preferred choice does not seem to be  available.
It would also be interesting to know whether $\mathfrak{C}_{\alpha\beta\gamma\delta}$ satisfies some hyperbolic system of evolution equations.
}
\end{remark}

Given a  smooth $(3+1)$-dimensional space-time $(\mcM,g)$, solution to Einstein's vacuum field equations with  cosmological constant $\Lambda$,  Theorem~\ref{int_main_result}  provides an algorithmic procedure to check whether  a given $\Lambda$-vacuum space-time
$(\mcM,g)$  admits a (possibly complex) KVF such that  the associated  MST vanishes.
The procedure is algorithmic in the sense that only differentiation and computation of roots are needed without any need to solve
differential equations:

The first step is to compute the  self-dual Weyl tensor $\mathcal{C}_{\mu\nu\sigma\rho}$ from $g_{\mu\nu}$.
We assume that the conditions \eq{all_ineq_C2}  hold.%
\footnote{
\label{footnote_codimension}
In case $\pm\sqrt{\mathcal{C}^2} - \sqrt{\frac{32}{3}}\,\Lambda\ne 0$ or $\pm\sqrt{\mathcal{C}^2} + \sqrt{\frac{8}{3}} \Lambda \ne 0$ are is violated on surfaces of co-dimension $\geq 1$, we restrict attention to  those regions of the space-time where both  are satisfied.
Once we know that the space-time belongs to e.g.\  the Kerr-NUT-(A)dS-family off these surfaces, it follows from smoothness that they are Kerr-NUT-(A)dS
everywhere.
It will be shown in Section~\ref{sect_scalar_invs} that Kerr-(A)dS satisfies $\mathcal{C}^2 \ne 0$, whence we can simply require  that
 $\mathcal{C}^2 \ne 0$ holds everywhere.
}
Then it merely remains to be checked whether the tensor $\mathfrak{C}_{\alpha\beta\gamma\delta}$ vanishes.
If and only if it  does, the MST vanishes for some non-trivial KVF $X$ which can be computed via \eq{final_eqn_KVF} in terms of
$\mathcal{C}_{\mu\nu\sigma\rho}$.

%

\subsection{An alternative algorithm}
\label{sec_alt}

One may combine \eq{main_cond} and \eq{Killing_cand2}  into one equation,
%
\begin{equation}
X_{\mu}X_{\nu}
\,=\, \pm
\Big(\frac{2}{3}\Big)^{1/2}\widetilde\varkappa^2
 (\mathcal{C}^2)^{-17/6}\Big(\mathcal{C}_{\mu\alpha\nu\beta}
\pm \sqrt{\frac{\mathcal{C}^2}6}\,\mathcal{I}_{\mu\alpha\nu\beta}\Big) \nabla^{\alpha} \mathcal{C}^2 \nabla^{\beta} \mathcal{C}^2
\;.
\label{eqn_in_terms_of_X}
\end{equation}
%
%
%
An argument analog to the one given in Section~\ref{sec_uniqueness}
 shows that a solution cannot exist simultaneously for both for $+$ and $-$.
%
%

A necessary condition for a $\Lambda$-vacuum space-time which fulfills the conditions  \eq{all_ineq_C2}
  to admit a  KVF w.r.t.\ which the MST vanishes is that there exists a $\varkappa\in \mathbb{C}\setminus\{0\}$ (which may be set equal to one)
 such that \eq{eqn_in_terms_of_X}
admits a  solution $X$. Consequently,  $X$
is  the only candidate  field with these properties up to multiplication with some non-vanishing constant
(there are two candidates as long as the ``right'' sign in \eq{eqn_in_terms_of_X} has not been figured out).
An explicit solution to \eq{eqn_in_terms_of_X} can be determined similarly to the methods described in Section~\ref{sec_solv}.

%


Let us explore some consequences of  \eq{eqn_in_terms_of_X}.
Taking the trace, we recover  \eq{norm_candidate_field},
%
%
while contraction with $X^{\nu}$ yields
\begin{equation}
X_{\mu}
\,=\, \pm
\Big(\frac{2}{3}\Big)^{1/2}
 \widetilde\varkappa^2|X|^{-2}  (\mathcal{C}^2)^{-17/6}\Big(X^{\nu}\mathcal{C}_{\mu\alpha\nu\beta}
\pm {\sqrt{\frac{\mathcal{C}^2}6}}X^{\nu}\mathcal{I}_{\mu\alpha\nu\beta}\Big) \nabla^{\alpha} \mathcal{C}^2 \nabla^{\beta} \mathcal{C}^2
,
\label{X_expression_X}
\end{equation}
where we need to assume  $|X|^2\ne 0$, or, equivalently (compare \eq{norm_candidate_field}), $|\nabla\mathcal{C}^2|^2\ne 0$.
We contract \eq{X_expression_X}  with $\nabla^{\mu}\mathcal{C}^2$ to deduce that
\begin{equation}
X^{\mu}\nabla_{\mu}\mathcal{C}^2
\,=\, 0
\;.
\label{X_XC}
\end{equation}
We insert \eq{norm_candidate_field}  and \eq{X_XC}  into \eq{X_expression_X},
\begin{eqnarray}
X_{\mu}
&=& 
\frac{2\sqrt{6}}{\sqrt{\mathcal{C}^2}
 \nabla_{\gamma} \mathcal{C}^2 \nabla^{\gamma} \mathcal{C}^2}\nabla^{\alpha} \mathcal{C}^2 \nabla^{\beta} \mathcal{C}^2\mathcal{C}_{\mu\alpha\nu\beta} X^{\nu}
\;.
\end{eqnarray}

\begin{lemma}
\label{lem_cand_fields}
Let $(\mcM,g)$ be a  smooth $(3+1)$-dimensional $\Lambda$-vacuum space-time  which satisfies $|\nabla\mathcal{C}^2|\ne 0$ and the inequalities \eq{all_ineq_C2}.
Consider the   2-tensor
$$
\mathcal{B}^{\mu}{}_{\nu}\,:=\,\frac{2\sqrt{6}}{\sqrt{\mathcal{C}^2}
 \nabla_{\gamma} \mathcal{C}^2 \nabla^{\gamma} \mathcal{C}^2}\nabla^{\alpha} \mathcal{C}^2 \nabla^{\beta} \mathcal{C}^2\mathcal{C}^{\mu}{}_{\alpha\nu\beta}
\;.
$$
A necessary condition for $(\mcM,g)$ to admit a (possibly complex)   KVF w.r.t.\ which the MST vanishes is that
$1$ is an eigenvalue of $\mathcal{B}$.
\end{lemma}

An analysis analog to the one in Section~\ref{sec_petrov} shows that  \eq{eqn_in_terms_of_X} admits a solution
if and only if
\begin{equation}
\mathcal{A}_{\mu\sigma}\mathcal{A}^{\sigma}{}_{\nu} \,=\, \tr\mathcal{A}\, \mathcal{A}_{\mu\nu}\;, \quad
\mathcal{A}_{\mu\nu}\,:=\,  \Big(\mathcal{C}_{\mu\alpha\nu\beta}
\pm  {\sqrt{\frac{\mathcal{C}^2}6}}\,\mathcal{I}_{\mu\alpha\nu\beta}\Big) \nabla^{\alpha} \mathcal{C}^2 \nabla^{\beta} \mathcal{C}^2
\;,
\end{equation}
or, equivalently (using \eq{weyl_rel1} and assuming $|\nabla\mathcal{C}^2|\ne 0$ ),
\begin{equation}
\mathfrak{C}_{\mu\alpha\nu\delta}
\nabla^{\alpha} \mathcal{C}^2  \nabla^{\delta} \mathcal{C}^2
\,=\,0
\;.
\label{trace_item_weyl}
\end{equation}
%

Finally, one needs to check that the so-obtained $X$ is indeed a KVF whose associated MST vanishes.
It might be possible to  show, in a similar manner to what we did in Section~\ref{sec_KVF_exist}, that \eq{eqn_in_terms_of_X}
 already implies this property.
Clearly, \eq{trace_item_weyl}  holds whenever \eq{lemma_C_tensor} holds.
In fact, these considerations  indicate that \eq{eqn_in_terms_of_X} may not suffice to ensure $\mathfrak{C}_{\alpha\beta\mu\nu}=0$,
which among other things  ensures that $(\mcM,g)$ is of Petrov type D.
So  one might expect
that additional conditions are needed to guarantee  existence of a KVF with vanishing MST.
 We have not  investigated this any further,
since  we are primarily interested in
 the algorithmic approach described in Section~\ref{section_alg_charact}.



\section{An algorithmic space-time characterization of the Kerr-NUT-(A)dS family}
\subsection{Vanishing of the MST associated to real KVFs}
\label{sec_van_real_KVF}

We want to provide an algorithmic characterization for the Kerr-NUT-(A)dS metrics.
A necessary condition for them is that the space-time admits a \emph{real} KVF such that the associated  MST vanishes.
While we did not care  so far whether the KVF for which the associated MST vanishes was real or complex, we henceforth
need to require the existence of a \emph{real} KVF with this property.

First of all, a  remark is in order concerning the ``gauge constant'' $\widetilde\varkappa$. As long as it does not matter whether  KVF is real, $\widetilde\varkappa$ can  simply be set
equal to $1$.
Now, we need to make sure  the existence of  a non-zero constant $\widetilde\varkappa$ such that
the vector field $X$, as given by
\eq{final_eqn_KVF},  is real.

Only the absolute value of $\widetilde\varkappa$
 remains as gauge freedom.
Together with the freedom to choose the sign of $\mathcal{F}_{\alpha\beta}$ this
reflects the fact that $X$ is a real KVF w.r.t.\ which the MST vanishes if and only if $\mu X$ has this property for some
$\mu \in \mathbb{R}\setminus \{0\}$.


\begin{proposition}
\label{prop_real_KVF}
Let $(\mcM,g)$ be a
smooth $(3+1)$-dimensional $\Lambda$-vacuum space-time such that
(cf.\ footnote~\ref{footnote_weaker})
\begin{equation}
\mathcal{C}^2 \,\ne\, 0\;, \quad   \mathcal{C}^2-\frac{32}{3}\Lambda^2\,\ne\, 0\;,
\quad  \mathcal{C}^2- \frac{8}{3} \Lambda^2 \,\ne\, 0
\;.
\label{all_ineq_C_b}
\end{equation}
%
Then,  if and only if $\mathfrak{C}_{\alpha\beta\gamma\delta}=0$
and
there exists a complex constant
\begin{equation}
\widetilde\varkappa
\in\mathbb{C}\setminus\{0\}
\;,
\end{equation}
 for which
\begin{equation}
 X^{\alpha} \,=\,
\widetilde\varkappa (\mathcal{C}^2)^{-1/6}
\Big(\mathcal{C}^{\alpha\beta}{}_{\mu\nu} \mathcal{W}^{\mu\nu}
\pm\sqrt{ \frac{\mathcal{C}^2}{6}}\,\mathcal{W}^{\alpha\beta}\Big) \nabla_{\beta}\log \mathcal{C}^2
\label{final_eqn_KVF2}
\end{equation}
is real for some self-dual tensor $\mathcal{W}_{\mu\nu}$  with
$|  \mathcal{C}^{\alpha\beta}{}_{\mu\nu} \mathcal{W}^{\mu\nu}
\pm\sqrt{ \frac{\mathcal{C}^2}{6}}\,\mathcal{W}^{\alpha\beta}|^2=1$,
there exists (up to rescaling) a unique  non-trivial real KVF  such that the associated MST vanishes. It is  given by
\eq{final_eqn_KVF2}.
\end{proposition}

Once a choice  of the branch of the roots has been made (cf.\ the discussion on p.\ \pageref{discussion_root}),
there exists, modulo $2\pi$, at most one argument of the complex number  $\widetilde \varkappa$ for which $X$ is real.

\subsection{Kerr-NUT-(A)dS family}
\label{sec_NUT}

It is well-known that the vanishing of the MST
associated to a \emph{real} KVF
 together
with the $\Lambda$-vacuum Einstein equations does not suffice to deduce that a given space-time $(\mcM,g)$ is locally isometric to a member of the
Kerr-NUT-(A)dS family. We will therefore fall back upon the characterization results reviewed in Section~\ref{sec_results}.
Once it is known that  $(\mcM,g)$ admits a real  KVF w.r.t.\ which the MST vanishes,
Theorem~\ref{thm_charact_aux}, or Theorem~\ref{thm_charact1} for $\Lambda=0$,
 can be consulted to check whether the emerging space-time belongs to the Kerr-NUT-(A)dS family.

Moreover, Theorem~\ref{thm_charact_aux}  can be used to express  the Kerr-NUT-(A)dS parameters $m$, $a$ and $\ell$ in terms of the constants $b_1$, $b_2$, $c$, and $k$, i.e.\ in terms of $g$ (and derivatives thereof), so  that one gains insights which member of the family  one has been given.
For that purpose recall the definitions \eq{equation_b1b2}-\eq{equation_k}
of the functions $b_1$, $b_2$, $c$, and $k$.
Using  $Q=Q_{\mathcal{F}}$, \eq{yet_another_Q}, \eq{PDE_Q2}, \eq{cand_F2} and \eq{norm_candidate_field} we find
%
\begin{eqnarray}
b_1 &=& 
3 \Big(\frac{3}{2}\Big)^{3/2}\,\mathrm{Im} ( \widetilde\varkappa^{3})
\,,
\label{gen_b1}
\\
 b_2  &=&  
-3\Big(\frac{3}{2}\Big)^{3/2}\,\mathrm{Re} (  \widetilde\varkappa^{3})
\,,
\\
c
&=&
-  \frac{\widetilde\varkappa^2}{4}
 (\mathcal{C}^2)^{-7/3}\nabla^{\alpha} \mathcal{C}^2\nabla_{\alpha} \mathcal{C}^2
\mp   \Big(\frac{3}{2}\Big)^{3/2}\mathrm{Re}[ \widetilde\varkappa^2 (\mathcal{C}^2)^{1/6} ]
\nonumber
\\
&&
-  3 \Lambda \,\mathrm{Re}[\widetilde\varkappa^2( \mathcal{C}^2)^{-1/3}]
\,,
\label{c_KdS}
\\
k
&=&
 9 \big| \widetilde\varkappa^2(\mathcal{C}^2)^{-1/3} \big| \nabla_{\mu}Z\nabla^{\mu}Z -  b_2Z +cZ^2 +\frac{\Lambda}{3} Z^4
\,,
\label{k_KdS}
\end{eqnarray}
where
\begin{eqnarray}
 Z  &\equiv&  6\,\mathrm{Re} \Big( \frac{\sqrt{\mathcal{F}^2}}{Q\mathcal{F}^2-4\Lambda}\Big)
\,=\,  3\,\mathrm{Re}[\widetilde\varkappa(\mathcal{C}^2)^{-1/6}]
\,,
\\
\nabla_{\mu}Z
 &=& -\frac{1}{2}\mathrm{Re}\Big(  \widetilde\varkappa(\mathcal{C}^2)^{-7/6}
\nabla_{\mu}\mathcal{C}^2
\Big)
\;.
\label{gen_nablaZ}
\end{eqnarray}
%
%


Moreover,
\begin{eqnarray}
 y &:=&  6\,\mathrm{Im} \Big( \frac{\sqrt{\mathcal{F}^2}}{Q\mathcal{F}^2-4\Lambda}\Big)
=
3\,\mathrm{Im} [ \widetilde\varkappa(\mathcal{C}^2)^{-1/6}]
\,.
\end{eqnarray}

\begin{remark}
\label{remark_weaker}
{\rm
The 6th root $(\mathcal{C}^2)^{1/6}$ is determined by the requirement that the Killing vector field $X$ needs to be real, cf.\ \eq{final_eqn_KVF2}.
}
\end{remark}

\subsection{Kerr-(A)dS family}
\label{sec_Kerr_AdS}
\subsubsection{An algorithmic characterization of the  Kerr-(A)dS  family}

Since it is of particular physical interest and
the algorithm can be presented in a more explicit  way,
 we devote attention henceforth   to the Kerr-(A)dS family.
To give an algorithmic local characterization result for Kerr-(A)dS
we   rely on  Corollary~\ref{cor_charact_aux}.
Assume we have been given a
$\Lambda$-vacuum space-time $(\mcM,g)$ which satisfies the inequalities \eq{all_ineq_C_b}.
Assume further that the algorithm described in Proposition~\ref{prop_real_KVF} shows that $(\mcM,g)$ admits a real KVF $X$
 such  the associated MST vanishes,  and  where $\mathrm{grad}(y)$ is not identically zero.
Then $Q\mathcal{F}^2$ and $Q\mathcal{F}^2 -4 \Lambda$ are not identically zero, and
Corollary~\ref{cor_charact_aux} can be applied.


According to  this  corollary a necessary condition for $(\mcM,g)$ to be  locally isometric to a Kerr-(A)dS space-time is  $b_2=0$, or
$\mathrm{Re}(\widetilde\varkappa^3)=0$.
In fact, we may  employ the gauge freedom contained in $\widetilde\varkappa$,
  which arises from the freedom to rescale the KVF, to achieve e.g.\ that
%
\begin{equation}
\widetilde \varkappa 
\,=\, i
\;.
\label{gauge_varkappa}
\end{equation}
%

%


\subsubsection{Scalar invariants of the Kerr-(A)dS family}
\label{sect_scalar_invs}


We want to compute $\mathcal{C}^2$ for the Kerr-(A)dS family.
In space-times  with vanishing MST and $\mathcal{F}^2\ne 0$, it is  convenient to do that via $\mathcal{F}^2$.
It follows from \eq{rln_Q2F4_C2} and \eq{intermediate_rln}
that
\begin{eqnarray}
\mathcal{C}^2 &=& \frac{2}{3}Q^2\mathcal{F}^4\,=\,  \frac{3}{2}\mathcal{F}^{-4}( \mathcal{F}^{\alpha\beta} \mathcal{F}^{\mu\nu}\mathcal{C}_{\alpha\beta\mu\nu} )^2
\;.
\end{eqnarray}
We consider  the Kerr-(A)dS metric  in standard
Boyer-Lindquist-type coordinates,
%
\begin{eqnarray}
g &=&
- \frac{\Delta_r- a^2\Delta_{\theta}\sin^2\theta}{\Sigma \Xi^2}\mathrm{d}t^2
 +2 \frac{ \Delta_r  -  (a^2+r^2) \Delta_{\theta}}{\Sigma \Xi^2}a\sin^2\theta\mathrm{d}t \mathrm{d}\phi
\nonumber
\\
&&
+\frac{(a^2+r^2)^2 \Delta_{\theta}  - a^2\sin^2\theta \Delta_r}{\Sigma \Xi^2} \sin^2\theta\mathrm{d}\phi^2
+ \frac{\Sigma}{\Delta_r}\mathrm{d}r^2+\frac{\Sigma}{\Delta_{\theta}}\mathrm{d}\theta^2
\;,
\label{KdS_metric}
\end{eqnarray}
where
\begin{eqnarray}
\Sigma &:=& r^2 +a^2\cos^2\theta
\;,
\\
\Xi &:=& 1 +\frac{\Lambda}{3}a^2
\;,
\\
\Delta_{\theta} &:=& 1 + \frac{\Lambda}{3} a^2\cos^2\theta
\;,
\\
\Delta_r &:=& (a^2+r^2)\Big( 1-\frac{\Lambda}{3}r^2\Big) -2mr
\;,
\end{eqnarray}
and
\begin{equation}
t\in\mathbb{R}\;, \quad r\in\mathbb{R}\;, \quad \theta\in [0,\pi]\;, \quad \phi\in[0,2\pi)
\;.
\end{equation}
We find for $X=\partial_t$, and  away from the ring singularity $\{r=0,\theta=\pi/2\}$ for $a\ne 0$, and the spacelike  $r=0$-singularity for $a=0$,
\begin{eqnarray}
\mathcal{F}^2 \,=\, - 4\Big(
\frac{3 m  -\Lambda r(  r^2 - 3a^2  \cos^2\theta) - ia\Lambda\cos\theta (3  r^2   -  a^2   \cos^2\theta )}
{(3 +a^2\Lambda)(r + i a \cos\theta)^2}
\Big)^2
\;,
\\
\mathcal{F}^{\alpha\beta} \mathcal{F}^{\mu\nu}\mathcal{C}_{\alpha\beta\mu\nu}  \,=\,
-\frac{8m}{ (r + i a\cos\theta)^3}  \mathcal{F}^2
\;.
\hspace{13em}
\end{eqnarray}
We conclude that $\mathcal{F}^2=0$ if and only if
\begin{equation}
3 m - \Lambda r\big( r^2 - 3a^2 \cos^2\theta\big)  \,=\,0
\quad \text{and}\quad
a\Lambda\cos\theta  \big( 3 r^2 -a^2  \cos^2\theta \big) \,=\,0
\;.
\label{F2_zero}
\end{equation}
Assume  $m\ne  0$. Then \eq{F2_zero} will never be satisfied for $\Lambda =0$.
For $\Lambda\ne 0$ they are satisfied, i.e.\ $\mathcal{F}^2=0$, if and only if
\begin{eqnarray}
\text{for $a=0$:} &&    r  \,=\,  \big(3 m\Lambda^{-1}\big)^{1/3}
\;,
\label{possi1}
\\
\text{for $a\ne   0$:} && \theta  \,=\,\pi/2
\quad \text{and} \quad
    r  \,=\,  \big(3 m\Lambda^{-1}\big)^{1/3}
\label{possi2}
\\
&& \hspace{5em}  \text{or}
\nonumber
\\
&&
  \cos\theta \,=\, \pm \Big(\frac{9}{8}\sqrt{3}\, ma^{-3} \Lambda^{-1}\Big)^{1/3}
\quad \text{and} \quad
  r   \,=\,  \mp \frac{a}{\sqrt{3}}  \cos\theta
\;.
\label{possi3}
\end{eqnarray}
Clearly,  the solution \eq{possi3} exists only for $\frac{9}{8}\sqrt{3}\, ma^{-3} \Lambda^{-1}\leq 1$.

We deduce that
\begin{eqnarray}
\mathcal{C}^2 \,=\,  \frac{96m^2}{ (r + i a\cos\theta)^6}
\;,
\end{eqnarray}
%
%
which by continuity remains true on the surfaces \eq{possi1}-\eq{possi3}.
For $m\ne 0$ it  is   non-zero everywhere. As one should expect $\mathcal{C}^2$ diverges  when approaching  the ring singularity for $a\ne 0$ and the spacelike $r=0$-singularity for $a=0$.


The tensor $\mathfrak{C}_{\alpha\beta\gamma\delta}$, cf.\ \eq{item_weyl_0}, vanishes for ``$-$''  if we  define $\sqrt{\,\cdot\,}$ such that
\begin{equation}
\sqrt{\mathcal{C}^2} \,=\, \frac{\sqrt{96} \,m}{ (r + i a\cos\theta)^3}
\;.
\end{equation}
%
%
Let us  analyze the remaining conditions in \eq{all_ineq_C_b} (cf.\ footnote~\ref{footnote_weaker}), which we need to impose for $\Lambda \ne 0$,
\begin{eqnarray}
  \sqrt{\mathcal{C}^2} \,=\, \sqrt{\frac{32}{3}}\,\Lambda & \Longleftrightarrow  &
3 m\Lambda^{-1}  \,=\,  (r + i a\cos\theta)^3
\;,
\label{ineq_eq1}
\\
\quad \sqrt{ \mathcal{C}^2} \,=\,- \sqrt{\frac{8}{3}} \Lambda  & \Longleftrightarrow  &
6m\Lambda^{-1}  \,=\,  - (r + i a\cos\theta)^3
\;.
\label{ineq_eq2}
\end{eqnarray}
Now
\begin{eqnarray}
\mathrm{Re}[ (r + i a\cos\theta)^3] &=& r(r^2-3a^2\cos^2\theta)
\;,
\\
 \mathrm{Im}[ (r + i a\cos\theta)^3] &=& a\cos\theta(\sqrt{3}\, r - a\cos\theta)(\sqrt{3} r + a\cos\theta)
\;.
\end{eqnarray}
One finds that \eq{ineq_eq1} holds if and only if $\mathcal{F}^2=0$, i.e.\ if and only if \eq{possi1}-\eq{possi3} hold.
Moreover,  \eq{ineq_eq2} is equivalent to
\begin{eqnarray}
\text{for $a=0$ :} &&    r  \,=\, - \big(6 m\Lambda^{-1}\big)^{1/3}
\;,
\label{possiB1}
\\
\text{for $a>  0$ :} && \theta  \,=\,\pi/2
\quad \text{and} \quad
    r  \,=\,  -\big(6 m\Lambda^{-1}\big)^{1/3}
\label{possiB2}
\\
&& \hspace{5em}  \text{or}
\nonumber
\\
&&
  \cos\theta \,=\, \pm \Big(\frac{9}{4}\sqrt{3}\, ma^{-3} \Lambda^{-1}\Big)^{1/3}
\quad \text{and} \quad
  r   \,=\,  \pm \frac{a}{\sqrt{3}}  \cos\theta
\;, \phantom{xxx}
\label{possiB3}
\end{eqnarray}
where \eq{possiB3} appears  only for $\frac{9}{4}\sqrt{3}\, ma^{-3} \Lambda^{-1}\leq 1$.%
%
%
\footnote{
At first glance this result seems to be in contradiction with  Proposition \ref{constancy}, which states that
if $Q\mathcal{F}^2$and $Q\mathcal{F}^2-\Lambda$ are not identically zero, which is the case for Kerr-(A)dS, then $\mathcal{F}^2$
is
 nowhere vanishing. We have just  seen that this is not true for Kerr-(A)dS.
The resolution simply is that the function $Q$ is assumed to be regular in  Proposition~\ref{constancy}, while it diverges for Kerr-(A)dS
at those points where $\mathcal{F}^2$ has zeros.
}

To sum it up,   for $\Lambda=0$, the conditions \eq{all_ineq_C_b} are satisfied everywhere by the Kerr family.
For $\Lambda\ne 0 $ and $a=0$, i.e.\ in the Schwarzschild-(A)dS case, they are violated on certain 3-surfaces, while
for  $\Lambda\ne 0 $ and $a>0$, i.e.\ in the proper Kerr-(A)dS case they are violated on certain 2-surfaces $\{r,\theta=\mathrm{const.}\}$.

Combining these   results  with the previous ones
we are led to  our second main result.

\begin{theorem}
\label{thm_main_result}
\label{thm_main_result2}
Let  $(\mcM, g)$ be a smooth
non-maximally symmetric%
\footnote{To exclude the Minkowski/(A)de Sitter-case where $m=0$ and $\mathcal{C}^2=0$.}
$(3+1)$-dim.\ space-time which satisfies  Einstein's vacuum field equations
with cosmological constant $\Lambda$. Then,  the space-time $(\mcM, g)$ is locally isometric to a member of the Kerr-(A)dS family  if and only if
\begin{enumerate}
\item[(i)] $\mathcal{C}^2\ne 0$ everywhere,
\item[(ii)] $\mathfrak{C}_{\alpha\beta\gamma\delta}\,\equiv \, \mathcal{C}_{\alpha\beta\mu\nu} \mathcal{C}^{\mu\nu}{}_{\gamma\delta}
\mp\sqrt{\frac{\mathcal{C}^2}{6}}\,\mathcal{C}_{\alpha\beta\gamma\delta}
-
\frac{\mathcal{C}^2}{3}\,\mathcal{I}_{\alpha\beta\gamma\delta}= 0 $,%
\footnote{For $\mathcal{C}^2\ne 0$ a solution exists at most for either $+$ or $-$.}
(equivalently, \eq{equiv_cond1}-\eq{equiv_cond2} hold),
\item[(iii)]
$\mathcal{C}^2 \ne \frac{32}{3}\,\Lambda^2$, and $\mathcal{C}^2\ne \frac{8}{3} \Lambda^2 $
(cf.\ footnote~\ref{footnote_weaker}),
\item[(iv)] $
 \mathrm{Re}\Big[(\mathcal{C}^2)^{-7/6}
\Big(\mathcal{C}^{\alpha\beta}{}_{\mu\nu} \mathcal{W}^{\mu\nu}
\pm \sqrt{ \frac{\mathcal{C}^2}{6}}\,\mathcal{W}^{\alpha\beta}\Big) \nabla_{\beta} \mathcal{C}^2
\Big]=0
$
 for some
(and then any)  self-dual 2-tensor   $\mathcal{W}_{\mu\nu}$ which satisfies $| \mathcal{C}^{\alpha\beta}{}_{\mu\nu} \mathcal{W}^{\mu\nu}
\pm\sqrt{ \frac{\mathcal{C}^2}{6}}\,\mathcal{W}^{\alpha\beta}|^2 =1$,
\item[(v)] $\mathrm{grad}(\mathrm{Re} [ (\mathcal{C}^2)^{-1/6}])$ is not identically zero, and
\item[(vi)] the constants  $c$ and $k$,
 given in terms of $\mathcal{C}^2$ by
\eq{c_KdS}-\eq{k_KdS} (and \eq{gauge_varkappa}),
 satisfy, depending on the sign of the cosmological constant, \eq{second_condition1}-\eq{second_condition3}, respectively.
\end{enumerate}
The Kerr-(A)dS family has parameters
\begin{equation*}
m\,=\, \Big(\frac{3}{2}\Big)^{5/2}\Big(\frac{\Lambda}{3} \zeta_1^2 + c\Big)^{-3/2}\,, \quad a\,=\, \zeta_1\Big(\frac{\Lambda}{3} \zeta_1^2 + c\Big)^{-1/2}\,,
\end{equation*}
where $\zeta_1$ is given by \eq{first_condition1b}-\eq{first_condition3b}.
The KVF whose associated MST vanishes (in Boyer-Lindquist-type coordinates this is a multiple of the $\partial_t$-KVF)
satisfies
\begin{equation*}
 X^{\alpha} \,=\, \mathrm{Im}\Big[ (\mathcal{C}^2)^{-7/6}
\Big(\mathcal{C}^{\alpha\beta}{}_{\mu\nu} \mathcal{W}^{\mu\nu}
\pm \sqrt{ \frac{\mathcal{C}^2}{6}}\,\mathcal{W}^{\alpha\beta}\Big) \nabla_{\beta} \mathcal{C}^2
\Big]
\;.
\end{equation*}
\end{theorem}

Comparing this, for $\Lambda=0$ with the algorithmic Kerr characterization obtained in \cite[Theorem~2]{fs} (cf.\ also \cite[Theorem~3]{lobo}),
it seems that the conditions of Theorem~\ref{thm_main_result2} are  verifiable with less computational effort
(the characterization in \cite{bk} since it relies on asymptotic conditions).

\vspace{1.2em}
\noindent {\textbf {Acknowledgements}}
I am grateful to  Piotr  Chru\'sciel and Marc Mars for several valuable  comments and suggestions on a first draft of this manuscript.
Useful comments by Lars Andersson are thankfully acknowledged, as well.
The research  was funded by the Austrian Science Fund (FWF):  P 23719-N16 and P 28495-N27.

\end{document}